\documentclass[10pt,journal,final]{IEEEtran}
\usepackage{amsmath,amsfonts}
\usepackage{amsthm}
\usepackage{amssymb}
\usepackage{algorithmic}
\usepackage{algorithm}
\usepackage{array}
\usepackage{mathrsfs}
\usepackage[caption=false,font=footnotesize,labelfont=rm,textfont=rm]{subfig}
\usepackage{textcomp}
\usepackage{stfloats}
\usepackage{url}
\usepackage{mathtools}
\usepackage{verbatim}
\usepackage{graphicx}
\usepackage{cite}
\usepackage{autobreak}
\usepackage{nicematrix}
\usepackage{arydshln}
\usepackage{latexsym}
\usepackage{booktabs}
\usepackage{subeqnarray}
\usepackage{cases}
\usepackage{bbm}
\usepackage{float}
\usepackage{nomencl}
\usepackage{alphalph,etoolbox}
\usepackage{xcolor} 
\definecolor{commentcolor}{gray}{0.6} 
\patchcmd{\subequations}{\alph{equation}}{\alphalph{\value{equation}}}{}{}
\makenomenclature
\allowdisplaybreaks[4]

\usepackage{etoolbox}
\renewcommand\nomgroup[1]{%
	\item[\bfseries
	\ifstrequal{#1}{A}{Abbreviations}{%
	\ifstrequal{#1}{B}{Indices and Sets}{%
	\ifstrequal{#1}{C}{Parameters}{%
	\ifstrequal{#1}{D}{Variables}{%
	\ifstrequal{#1}{E}{Functions}{}}}}}%
]}
\newtheorem{lemma}{Lemma}
\newcommand{\E}{\mathbb{E}}

\begin{document}
\title{Adaptive Federated Learning to Optimize Integrated Flows in Cyber–Physical Data Centers}

\author{Junhong~Liu,~\IEEEmembership{Member,~IEEE}, Lanxin Du, Yujia Li,~\IEEEmembership{Member,~IEEE}, Rong-Peng Liu,~\IEEEmembership{Member,~IEEE},\\ 
Yunfeng Li,~\IEEEmembership{Member,~IEEE}, Fei Teng,~\IEEEmembership{Senior Member,~IEEE}, Francis Yunhe Hou,~\IEEEmembership{Fellow,~IEEE}
\thanks{This work was supported in part by the National Key R\&D Program of China under Grant 2023YFA1011301; in part by Guangdong Basic and Applied Basic Research Foundation under Grant 2023A1515011942; and in part by the Fonds de recherche du Qu\'ebec-secteur Nature et technologies (FRQ-Secteur NT) under Grant FRQ-Secteur NT 367013. \textit{(Corresponding author: Francis Yunhe Hou.)}}
\thanks{Junhong Liu is with the Rausser College of Natural Resources, University of California at Berkeley, CA 94720, USA (e-mail: junhongliu@berkeley.edu).}
\thanks{Lanxin Du and Fei Teng are with the Department of Electrical and Electronic Engineering, Imperial College London, London, UK (e-mail: dulanxin.max@gmail.com, f.teng@imperial.ac.uk).}
\thanks{Yujia Li is with the Energy Storage \& Distributed Resources Division, Lawrence Berkeley National Laboratory, Berkeley, CA 94720, USA (e-mail: yujiali@lbl.gov)}
\thanks{Yunfeng Li is with the Shenzhen International Graduate School, Tsinghua University, Shenzhen 518055, China (e-mail: yunfengli@ucsb.edu).}
\thanks{Rong-Peng Liu is with the Department of Electrical and Computer Engineering, McGill University, Montreal, QC H3A 0E9, Canada (e-mail: rpliu@eee.hku.hk).}
\thanks{Francis Yunhe Hou is with the Department of Electrical and Electronic Engineering, The University of Hong Kong, Hong Kong SAR, China (e-mail: {yhhou}@eee.hku.hk).} \vspace{-2.0em}}

\markboth{}%
{Liu \MakeLowercase{\textit{et al.}}: Adaptive Federated Learning to Optimize Integrated Flows in Cyber--Physical Data Centers}
\maketitle
\begin{abstract}
Data centers play an increasingly critical role in societal digitalization, yet their rapidly growing energy demand poses significant challenges for sustainable operation. To enhance the energy efficiency of geographically distributed data centers, this paper formulates a multi-period optimization model that captures the interdependence of electricity, heat, and data flows. The optimization of such integrated multi-domain flows inherently involves mixed-integer formulations and access to proprietary or sensitive datasets, which correspondingly exacerbate computational complexity and raise data-privacy concerns. To address these challenges, an adaptive federated learning-to-optimization approach is proposed, accounting for the heterogeneity of datasets across distributed data centers. To safeguard privacy, cryptography techniques are leveraged in both the learning and optimization processes. A model acceptance criterion with convergence guarantee is developed to improve learning performance and filter out potentially contaminated data, while a verifiable double aggregation mechanism is further proposed to simultaneously ensure privacy and integrity of shared data during optimization. Theoretical analysis and numerical simulations demonstrate that the proposed approach preserves the privacy and integrity of shared data, achieves near-optimal performance, and exhibits high computational efficiency, making it suitable for large-scale data center optimization under privacy constraints.

\end{abstract}



\mbox{}
\nomenclature[A]{\(Enc\)}{Encryption}
\nomenclature[A]{\(Cmp\)}{Computation}
\nomenclature[A]{\(PK/SK\)}{Public/Private key}
\nomenclature[A]{\(\text{Mdn}/\text{Slp}/\text{Std}\)}{Median/Slope/Standard deviation}

\nomenclature[B]{\(i,g,s\)}{Index of data center/generator/battery}
\nomenclature[B]{\(t,r\)}{Index of timepoint/learning round}
\nomenclature[B]{\(N\)}{Number of ensemble neural networks}
\nomenclature[B]{\({\cal I}/{\cal G}/{\cal S}/{\cal T}\)}{Set of data center/generator/battery/operation time}
\nomenclature[B]{\({\mathbf X}/{\mathbf{X}^{\dag}}\)}{Set of decision variables}
\nomenclature[B]{\({\mathbb R}/{\mathbb Z}\)}{Set of real numbers/integers}

\nomenclature[C]{\(\lambda_{t}^{\text{imp}}/\lambda_{t}^{\text{exp}}/\lambda_{t}^{\text{reg}}\)}{Prices of imported electricity/exported electricity/regulation service at time $t$}
\nomenclature[C]{\(\Delta t\)}{Length of each operation time interval}
\nomenclature[C]{\(u_{i,t}/L_{t}\)}{Workload arrival rate of data center $i$/Base electric load, at time $t$}
\nomenclature[C]{\(\alpha_{g}/\beta_{g}\)}{Coefficients of utility function for generator $g$}
\nomenclature[C]{\(c_{s}^{\text{deg}}/C_{i}^{\text{penalty}}\)}{Degradation cost of charging or discharging for battery $s$/Penalty for SLA violation}
\nomenclature[C]{\(R_{i}^{\text{max}}/R_{i}^{\text{nominal}}\)}{Max/Nominal processing capacity of data center $i$}
\nomenclature[C]{\(\eta_{i}^{\text{min}}/\eta_{i}^{\text{max}}\)}{Min/Max processing efficiency of data center $i$}
\nomenclature[C]{\(\overline{P_{i}^{\text{dyn}}}\)}{Max dynamic power of data center $i$}
\nomenclature[C]{\(P_{i}^{\text{idle}}\)}{Idle power consumption of data center $i$}
\nomenclature[C]{\(x_i^{max}\)}{Max number of servers of data center $i$}
\nomenclature[C]{\(\text{SLA}_{i}\)}{Service level agreement target of data center $i$}
\nomenclature[C]{\(P_{t}^{\text{imp,max}}/P_{t}^{\text{exp,max}}/P_{t}^{\text{contract}}\)}{Upper bounds of imported/exported/contracted electricity}
\nomenclature[C]{\(\beta_{i}^{\text{cool}}/\beta_{i}^{\text{heat}}/\beta_{i}^{\text{loss}}/C_{i}^{\text{dis}}\)}{Cooling effectiveness factor/Heat generation factor/Heat loss coefficient/Thermal capacitance of data center $i$}
\nomenclature[C]{\(P_{i}^{\text{cool,base}}/Q_{i}^{\text{cool,max}}\)}{Base cooling power/Max cooling capacity}
\nomenclature[C]{\(\text{COP}_{i}\)}{Coefficient of performance for cooling system of data center $i$}
\nomenclature[C]{\(T^{\text{ambient}}/T_{i}^{\text{min}}/T_{i}^{\text{max}}\)}{Ambient temperature/Min/Max operating temperature for data center $i$}
\nomenclature[C]{\(R_{g}^{\text{up}}/R_{g}^{\text{down}}\)}{Upper bounds of ramping up/down for generator $g$}
\nomenclature[C]{\(P_{g}^{\text{min}}/P_{g}^{\text{max}}\)}{Lower/Upper bounds of generation for generator $g$}
\nomenclature[C]{\(P_{s}^{\text{ch,max}}/P_{s}^{\text{dch,max}}\)}{Upper bounds of charging/discharging for battery $s$}
\nomenclature[C]{\(E_{s}^{\text{min}}/E_{s}^{\text{max}}\)}{Min/Max bounds of SoC for battery $s$}
\nomenclature[C]{\(\eta_{s}^{\text{ch}}/\eta_{s}^{\text{dch}}\)}{Charging/Discharging coefficients for battery $s$}

\nomenclature[D]{\(q_{i,t}/r_{i,t}/\delta_{i,t}/T_{i,t}\)}{Queue length/Processing capacity/SLA violation/Temperature, for data center $i$ at time $t$}
\nomenclature[D]{\(x_{i,t}/z_{s,t}/\delta_{t}^{\text{imp}}\)}{Number of operating servers for data center $i$/Charging mode of battery $s$/Trading mode with the wholesale market, at time $t$}
\nomenclature[D]{\(\eta_{i,t}^{\text{eff}}\)}{Processing efficiency of single server for data center $i$ at time $t$}
\nomenclature[D]{\(r_{i,t}^{\text{eff}}\)}{Total processed data for data center $i$ at time $t$}
\nomenclature[D]{\(p_{i,t}^{\text{dyn}}/p_{i,t}^{\text{server}}/P_{i,t}^{\text{cool}}/p_{i,t}^{\text{dc}}\)}{Dynamic/Computing/Cooling/Total electricity consumption, for data center $i$ at time $t$}
\nomenclature[D]{\(P_{t}^{\text{imp}}/P_{t}^{\text{exp}}\)}{Imported/Exported electricity from/to the wholesale electricity market}
\nomenclature[D]{\(p_{g,t}\)}{Electricity from generator $g$ at time $t$}
\nomenclature[D]{\(Q_{i,t}^{dis}/Q_{i,t}^{loss}/Q_{i,t}^{\text{cool}}\)}{Total dissipated thermal power/Thermal power exchanged with environment/Thermal power removed by cooling, for data center $i$ at time $t$}
\nomenclature[D]{\(t_{i,t}^{\text{queue}}\)}{Time required for processing existing queue data for data center $i$ at time $t$}
\nomenclature[D]{\(P_{s,t}^{\text{ch}}/P_{s,t}^{\text{dch}}/E_{s,t}\)}{Charging power/Discharging power/SoC of battery $s$ at time $t$}

\nomenclature[E]{\(\nabla\)}{First-order partial differentiation}
\nomenclature[E]{\(\E\)}{Expectation operator}
\nomenclature[E]{\(Clip/abs\)}{Clipping/Absolute value operators}
\printnomenclature[2.2cm]






\section{Introduction}
\IEEEPARstart{T}{he} rapid proliferation of artificial intelligence (AI) and large language models (LLMs) in smart cities is driving an unprecedented demand for data processing, storage, and analytics. This trend is accelerating the deployment of hyperscale data centers worldwide, whose electricity consumption is projected to rise from 1--2\% today to 3--8\% of global usage by 2030 \cite{zhang2023research, koot2021usage}. Their operations involve the complex interplay of power supply, cooling systems, and computational workloads, forming a multi-energy system that integrates electricity, heat, and information flows \cite{fan2025highly,yin2022exploiting}. Efficiently optimizing such systems is therefore critical for enhancing overall energy efficiency and alleviating the environmental footprint.

Conventional data center energy management often relies on the holistic centralized optimization that requires complete access to operational and load data from all facilities\cite{rong2016optimizing}. However, the growing geographic distribution of data centers and the increasing sensitivity of proprietary data pose several challenges for the centralized approach. First, centralized operation limits the computational scalability, as multi-period formulations with mixed integers become intractable for large networks \cite{long2023collaborative,11142335}. Meanwhile, this holistic operation requires collecting real-time data through cyber infrastructures from multiple stakeholders, i.e., energy providers, computing service users, and possibly communication network operators, which can raise privacy leakage concerns \cite{faquir2021cybersecurity, karale2021challenges}. Recent years have witnessed increasing incidences of data breaches from leading technology platforms, affecting millions of users and exposing sensitive operational data \cite{saraswat2022protecting}. In fear of illegal data collection and abuse, stakeholders in the computational ecosystem can be conservative in sharing operational data, e.g., workload characteristics and energy consumption patterns, potentially causing inefficient resource allocation and further degrading system performance. To guarantee the efficient operation of hyperscale data centers, it is therefore imperative to prioritize the privacy and security of stakeholders' data. These challenges underscore the need for distributed and privacy-preserving approaches that can achieve near-optimal coordination without explicit data sharing.


Recent advances in federated learning offer a promising paradigm for distributed model training while keeping local data private. Researchers have proposed customized federated learning approaches for industrial applications \cite{aminifar2024privacy, alhazmi2025federated, yang2022federated, ahmad2024lightweight, saad2025towards}. To enhance the privacy preservation during federated learning, cryptography is further leveraged \cite{mantey2024federated, rieyan2024advanced}. Nevertheless, existing federated learning implementations as in \cite{aminifar2024privacy} normally assume homogeneous data distributions, thereby limiting their applicability to heterogeneous and dynamic data center environments. In addition, directly embedding federated learning into end-to-end optimization frameworks as in \cite{alhazmi2025federated} can lead to infeasible or suboptimal solutions due to the lack of convergence guarantees and constraint awareness. Furthermore, current federated learning or distributed optimization approaches as in \cite{mantey2024federated, rieyan2024advanced, saad2025towards,liu2023privacy} predominantly emphasize privacy preservation while overlooking data integrity, which is a critical aspect for ensuring trustworthy decision-making in large-scale networked infrastructures.

In parallel, cryptography-integrated aggregation protocols have been developed for general-purpose federated learning. Masking-based secure aggregation \cite{bonawitz2017practical} and batched additively homomorphic encryption \cite{zhang2020batchcrypt} protect the confidentiality of individual model updates, while verifiable aggregation schemes such as VerifyNet \cite{xu2020verifynet} and VeriFL \cite{guo2021verifl} further allow clients to verify the correctness of the aggregation results returned by the server. However, these schemes keep the aggregation rule fixed rather than adapting it to heterogeneous local data; conversely, heterogeneity-aware federated optimization methods \cite{li2020federated, karimireddy2020scaffold} provide convergence guarantees but leave privacy and integrity unaddressed. Moreover, the verification objects of \cite{xu2020verifynet, guo2021verifl} are limited to the aggregated model updates during training, whereas the operational data shared in the downstream energy optimization remain unprotected. Table \ref{tab:comparison} compares the proposed approach with the aforementioned adaptive and cryptography-integrated federated learning approaches in terms of salient features, advantages, and limitations.

\begin{table*}[!t]
	\centering
	\caption{Comparison with Representative Adaptive and Cryptography-Integrated Federated Learning Approaches}
	\label{tab:comparison}
	\scriptsize
	\setlength{\tabcolsep}{4pt}
	\renewcommand{\arraystretch}{1.15}
	\begin{tabular}{p{3.6cm}cp{2.7cm}cccp{4.1cm}}
		\toprule
		Approach (application domain) & \shortstack[c]{Heterogeneity\\awareness} & Privacy technique & \shortstack[c]{Integrity\\verification} & \shortstack[c]{Convergence\\analysis} & \shortstack[c]{Optimization\\integration} & Key limitation \\
		\midrule
		Bonawitz \textit{et al.} \cite{bonawitz2017practical} (general FL) & $\times$ & Pairwise masking + secret sharing & $\times$ & $\times$ & $\times$ & Aggregation results are not verifiable \\
		BatchCrypt \cite{zhang2020batchcrypt} (cross-silo FL) & $\times$ & Batched additive HE & $\times$ & $\times$ & $\times$ & Quantization errors; no integrity guarantee \\
		VerifyNet \cite{xu2020verifynet} (general FL) & $\times$ & Double masking & \checkmark$^{\dag}$ & $\times$ & $\times$ & Heavy cryptographic overhead for verification \\
		VeriFL \cite{guo2021verifl} (general FL) & $\times$ & Masking-based secure aggregation & \checkmark$^{\dag}$ & $\times$ & $\times$ & No adaptation to heterogeneous data \\
		FedProx \cite{li2020federated} (general FL) & \checkmark & $\times$ & $\times$ & \checkmark & $\times$ & No privacy or integrity protection \\
		SCAFFOLD \cite{karimireddy2020scaffold} (general FL) & \checkmark & $\times$ & $\times$ & \checkmark & $\times$ & No privacy or integrity protection \\
		Aminifar \textit{et al.} \cite{aminifar2024privacy} (mobile health) & $\times$ & Privacy-preserving edge FL & $\times$ & $\times$ & $\times$ & Assumes homogeneous data distributions \\
		Alhazmi \textit{et al.} \cite{alhazmi2025federated} (data center demand response) & $\times$ & FL without raw data sharing & $\times$ & $\times$ & \checkmark & Lacks convergence guarantee and constraint awareness \\
		Ahmad \textit{et al.} \cite{ahmad2024lightweight} (IoT attack detection) & $\times$ & Lightweight mini-batch FL & $\times$ & $\times$ & $\times$ & Lacks convergence guarantee \\
		Mantey \textit{et al.} \cite{mantey2024federated} (medical IoT) & $\times$ & Homomorphic encryption & $\times$ & $\times$ & $\times$ & Privacy preserved without data integrity \\
		Rieyan \textit{et al.} \cite{rieyan2024advanced} (data fabric) & $\times$ & Homomorphic encryption & $\times$ & $\times$ & $\times$ & Privacy preserved without data integrity \\
		Saad \textit{et al.} \cite{saad2025towards} (container orchestration) & $\times$ & Privacy-preserving FL & $\times$ & $\times$ & $\times$ & Privacy preserved without data integrity \\
		Liu \textit{et al.} \cite{liu2023privacy} (P2P energy trading) & $\times$ & Hybrid secure computations & $\times$ & --- & \checkmark & Optimization only; no learning; integrity overlooked \\
		\textbf{Proposed} (data center multi-energy management) & \checkmark & Secret sharing + CKKS/CRT-Paillier & \checkmark$^{\ddag}$ & \checkmark & \checkmark & Detection limited to injections above tolerance $\psi$ \\
		\bottomrule
		\multicolumn{7}{l}{$^{\dag}$Verification restricted to the aggregated model updates during training. \quad $^{\ddag}$Verification of the operational data shared in the optimization stage.}
	\end{tabular}
\end{table*}

To address these challenges, this paper proposes an adaptive federated learning-to-optimization framework for the energy management of data centers. The proposed approach accounts for the heterogeneity of personal datasets by adaptively adjusting the aggregate model parameters according to the acceptance criteria. In addition, cryptography techniques are leveraged in both the learning and optimization stages to safeguard data privacy, while a verifiable double aggregation mechanism is proposed to ensure the integrity and correctness of shared data. The main contributions of this work are summarized as follows:
\begin{itemize}
\item{We develop a coordinated cyber-physical multi-energy management model for hyperscale data centers that simultaneously optimizes the interactions among power supply, cooling systems, and data flows.}
\item{To account for the data heterogeneity and privacy preservation, we propose a near-optimal adaptive federated learning-to-optimization approach for distributed data center energy management. Meanwhile, to further enhance training convergence and filter out potentially contaminated aggregate updates during the learning stage, we also develop a model parameter acceptance criterion with convergence guarantee.}
\item{To simultaneously safeguard privacy and integrity of shared data in the optimization process, we further propose a lightweight verifiable double aggregation mechanism based on cryptography techniques.}
\end{itemize}
The remainder of this paper is organized as follows: Section II introduces the energy management of hyperscale data centers and its reformulations. Section III proposes the adaptive federated learning-to-optimization approach. Section IV demonstrates the effectiveness and scalability of the proposed approach with case studies. Section V draws the conclusions.

\section{Formulations of the Energy Management for Hyperscale Data Centers}
\subsection{Centralized Optimization Formulation}
This paper considers the day-ahead energy management for hyperscale data centers, which is formulated as $\mathcal{P}_1$:
\begin{subequations}
\begin{align}
&\mathop{\min}_{\mathbf{X}} \quad  \sum_{t \in \mathcal{T}} \{ P_{t}^{\text{imp}} \cdot \lambda_{t}^{\text{imp}} - P_{t}^{\text{exp}} \cdot \lambda_{t}^{\text{exp}}  \notag +\sum_{i \in \mathcal{I}} \delta_{i,t} \cdot C_{i}^{\text{penalty}} \\ 
&+\sum_{g \in \mathcal{G}} (\alpha_{g} \cdot p_{g,t} + \beta_{g} \cdot p_{g,t}^2 ) -\sum_{s \in \mathcal{S}}  P_{s,t}^{\text{dch}} \cdot \lambda_{t}^{\text{reg}}  \notag \\
&+ \sum_{s \in \mathcal{S}} ( P_{s,t}^{\text{ch}} + P_{s,t}^{\text{dch}}) \cdot c_{s}^{\text{deg}}\} \cdot \Delta t \label{e1a} \\
&\mathbf{s.t.} \quad   q_{i,0} = 0,  \forall i \in \mathcal{I}, \quad E_{s,0} = E_{s}^{\text{min}}, \forall s \in \mathcal{S}, \label{e1b} \\
&q_{i,t+1} = q_{i,t} + u_{i,t} \cdot \Delta t - r_{i,t}^{\text{eff}} \cdot \Delta t, \forall i \in \mathcal{I}, t \in \mathcal{T} \label{e1c} \\ 
&r_{i,t}^{\text{eff}} = r_{i,t} \cdot \eta_{i,t}^{\text{eff}}, \forall i \in \mathcal{I}, t \in \mathcal{T} \label{e1d}\\
&0 \leq  r_{i,t} \leq x_{i,t} \cdot R_{i}^{\text{max}}, \forall i \in \mathcal{I}, t \in \mathcal{T} \label{e1e}\\
&\eta_{i}^{\text{min}} \leq \eta_{i,t}^{\text{eff}} \leq \eta_{i}^{\text{max}},  \forall i \in \mathcal{I}, t \in \mathcal{T} \label{e1f}\\
&p_{i,t}^{\text{dyn}} = \frac{r_{i,t}^{\text{eff}}}{x_i^{max}\cdot R_{i}^{\text{max}}} \cdot \overline{P_{i}^{\text{dyn}}}, \forall i \in \mathcal{I}, t \in \mathcal{T} \label{e1g}\\
&p_{i,t}^{\text{server}} = x_{i,t} \cdot P_{i}^{\text{idle}} + p_{i,t}^{\text{dyn}},  \forall i \in \mathcal{I}, t \in \mathcal{T}  \label{e1h}\\
& 0 \le x_{i,t} \le x_i^{max}, x_{i,t} \in \mathbb{Z}, \forall i \in \mathcal{I}, t \in \mathcal{T} \label{e1i}\\
&\delta_{i,t} \geq t_{i,t}^{\text{queue}} - \text{SLA}_{i}, \ \delta_{i,t} \ge 0, \forall i \in \mathcal{I}, t \in \mathcal{T}  \label{e1j}\\
&t_{i,t}^{\text{queue}} \geq \frac{q_{i,t}}{R_{i}^{\text{nominal}}}, \forall i \in \mathcal{I}, t \in \mathcal{T} \label{e1k}\\
&0 \leq P_{t}^{\text{imp}} \leq \delta_{t}^{\text{imp}} \cdot P_{t}^{\text{imp,max}},  \forall t \in \mathcal{T} \label{e1l}\\
&0 \leq  P_{t}^{\text{exp}} \leq (1 - \delta_{t}^{\text{imp}}) \cdot P_{t}^{\text{exp,max}}, \forall t \in \mathcal{T} \label{e1m}\\
&\delta_{t}^{\text{imp}}, z_{s,t} \in \{0,1\},  \forall s \in \mathcal{S}, \forall t \in \mathcal{T} \\
&P_{t}^{\text{imp}} \leq P_{t}^{\text{contract}}, \forall t \in \mathcal{T} \label{e1n}\\
&T_{i,t} = T_{i,t-1} + \frac{p_{i,t-1}^{\text{server}} \cdot \beta_{i}^{\text{heat}}-Q_{i,t-1}^{dis}}{C_{i}^{\text{dis}}}, \forall i \in \mathcal{I}, t \in \mathcal{T} \label{e1o}\\
&Q_{i,t-1}^{dis}=Q_{i,t-1}^{\text{cool}} \cdot \beta_{i}^{\text{cool}}+Q_{i,t-1}^{loss}, \forall i \in \mathcal{I}, t \in \mathcal{T} \label{e1p}\\
&Q_{i,t-1}^{loss}= (T_{i,t-1} - T^{\text{ambient}}) \cdot \beta_{i}^{\text{loss}}, \forall i \in \mathcal{I}, t \in \mathcal{T} \label{e1q}\\
&P_{i,t}^{\text{cool}} = x_{i,t} \cdot P_{i}^{\text{cool,base}} + \frac{Q_{i,t}^{\text{cool}}}{\text{COP}_{i}}, \forall i \in \mathcal{I}, t \in \mathcal{T} \label{e1r}\\
&Q_{i,t}^{\text{cool}} \leq Q_{i}^{\text{cool,max}}, \forall i \in \mathcal{I}, t \in \mathcal{T} \label{e1s}\\
&T_{i}^{\text{min}} \leq T_{i,t} \leq T_{i}^{\text{max}}, T_{i,0} = T^{\text{ambient}}, \forall i \in \mathcal{I}, t \in \mathcal{T} \label{e1t}\\
&p_{i,t}^{\text{dc}} = p_{i,t}^{\text{server}} + P_{i,t}^{\text{cool}}, \forall i \in \mathcal{I}, t \in \mathcal{T} \label{e1u}\\ 
&p_{g,t} - p_{g,t-1} \leq R_{g}^{\text{up}} \cdot \Delta t,  \forall g \in \mathcal{G}, t \in \mathcal{T} \label{e1v}\\
&p_{g,t-1} - p_{g,t} \leq R_{g}^{\text{down}} \cdot \Delta t,  \forall g \in \mathcal{G}, t \in \mathcal{T}  \label{e1w}\\
&P_{g}^{\text{min}} \leq p_{g,t} \leq   P_{g}^{\text{max}}, \forall g \in \mathcal{G}, t \in \mathcal{T} \label{e1x}\\
&0 \leq  P_{s,t}^{\text{ch}} \leq z_{s,t} \cdot P_{s}^{\text{ch,max}}, \forall s \in \mathcal{S}, t \in \mathcal{T} \label{e1y}\\
&0 \leq P_{s,t}^{\text{dch}} \leq (1 - z_{s,t}) \cdot P_{s}^{\text{dch,max}}, \forall s \in \mathcal{S}, t \in \mathcal{T} \label{e1z}\\
&E_{s}^{\text{min}} \le E_{s,t} \le E_{s}^{\text{max}}, \forall s \in \mathcal{S}, t \in \mathcal{T} \label{e1aa}\\
&E_{s,t} = E_{s,t-1} + \{\eta_{s}^{\text{ch}} \cdot P_{s,t-1}^{\text{ch}} - \frac{P_{s,t-1}^{\text{dch}}}{\eta_{s}^{\text{dch}}}\} \cdot \Delta t, \forall s \in \mathcal{S}, t \in \mathcal{T} \label{e1ab}\\
&\sum_{g \in \mathcal{G}} p_{g,t} + P_{t}^{\text{imp}} - P_{t}^{\text{exp}}+ \sum_{s \in \mathcal{S}} (P_{s,t}^{\text{dch}}-P_{s,t}^{\text{ch}}) \notag \\ 
&= \sum_{i \in \mathcal{I}} p_{i,t}^{\text{dc}}+L_{t},  \forall t \in \mathcal{T}, \label{e1ac}
\end{align}
\end{subequations}
where $\mathbf{X}$ consists of the decision variables including $r_{i,t}, \eta_{i,t}^{\text{eff}}, T_{i,t}, r_{i,t}^{\text{eff}}, p_{i,t}^{\text{dyn}}, p_{i,t}^{\text{server}}, P_{i,t}^{\text{cool}}, Q_{i,t}^{\text{cool}},\delta_{i,t}, P_{t}^{\text{imp}}, P_{t}^{\text{exp}}, p_{g,t}$, $z_{s,t}, x_{i,t}, p_{i,t}^{\text{dc}}, \delta_{t}^{\text{imp}}, Q_{i,t}^{dis}, Q_{i,t}^{loss}, q_{i,t}, t_{i,t}^{\text{queue}}, P_{s,t}^{\text{ch}}, P_{s,t}^{\text{dch}}, E_{s,t}$. The total costs in \eqref{e1a} consist of energy import and export costs, service delay penalty cost, generation cost,  ancillary service income, charging and discharging degradation costs for the battery. \eqref{e1b} refers to the initial length of data queue for data center $i$ and initial SoC for storage $s$. \eqref{e1c} refers to the data queue evolution of the data center $i$, where $u_{i,t}$ denotes the workload arrival rate of data center $i$ at time $t$. \eqref{e1d} refers to the effective processing rate of the data center $i$. \eqref{e1e} refers to the maximum processing rate of the data center $i$. \eqref{e1h} refers to the total server power consumption, which consists of the idle and dynamic power consumption. \eqref{e1j} refers to the data queue time and SLA (service level agreement) constraints. \eqref{e1k} represents the nominal queue time, where $R_{i}^{\text{nominal}}=R_{i}^{\text{max}} \cdot 0.8$. \eqref{e1o} refers to the temperature evolution of the data centers. \eqref{e1q} refers to the heat dissipated to the environment, where $\beta_{i}^{\text{loss}}$ denotes the heat loss coefficient. \eqref{e1r} refers to the heat absorbed by the cooling devices, where COP$_i$ stands for the coefficient of performance for data center $i$. \eqref{e1u} refers to total power consumption for individual data center. \eqref{e1v}-\eqref{e1x} refer to the generator ramping conditions. \eqref{e1y}-\eqref{e1ab} refer to the energy evolution of battery storage systems. \eqref{e1ac} refers to the total system power balance, where $L_{t}$ denotes the base electric load at time $t$. The formulated problem is a mixed-integer non-linear programming (MINLP) problem.

\begin{figure}[!hbp]
	\centering
    \includegraphics[width=0.9\linewidth]{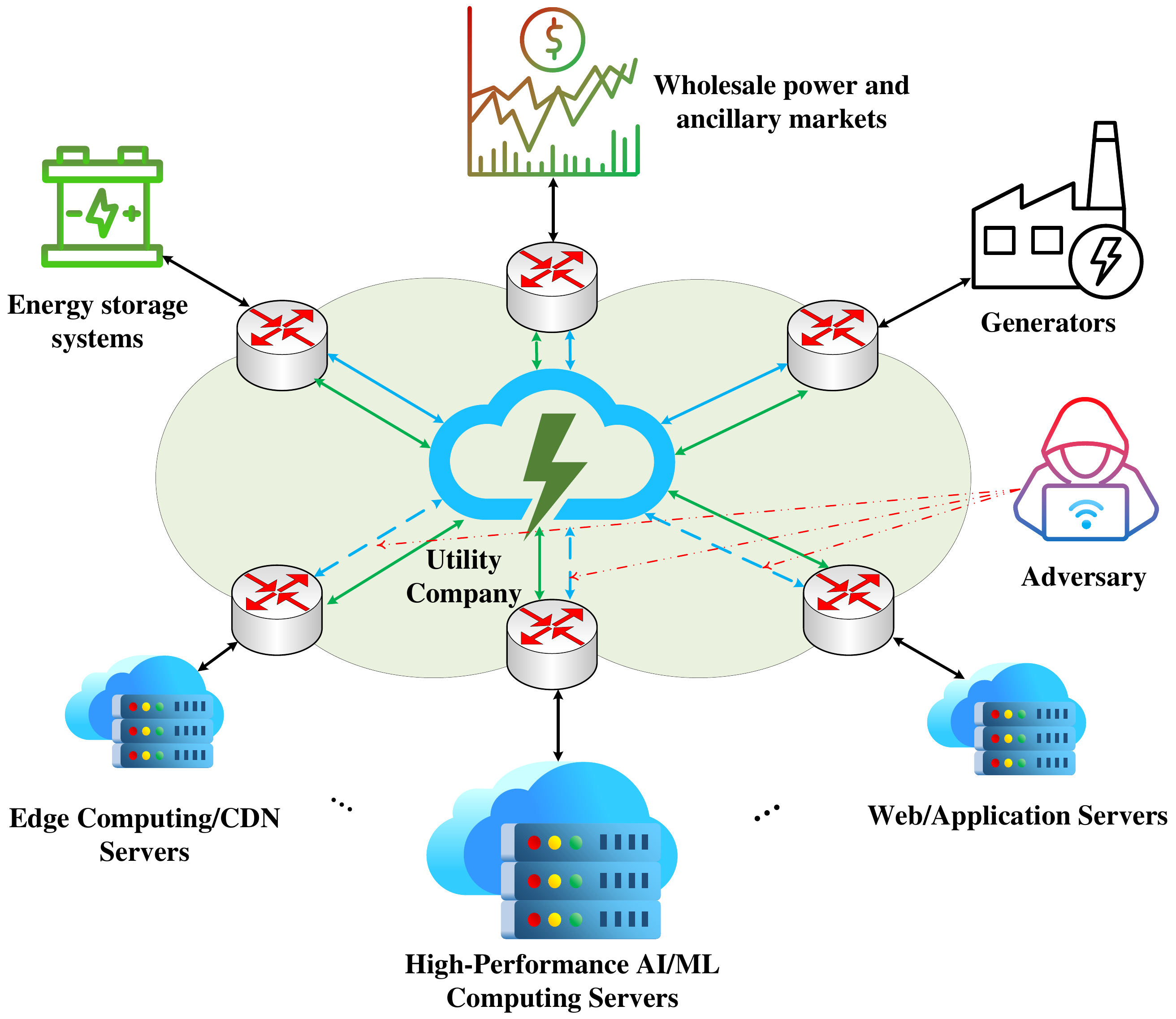}
	\caption{Scheme of the data center energy management.}
	\label{fig1}
\end{figure}

To handle the bilinear term $r_{i,t}^{\text{eff}} = r_{i,t} \cdot \eta_{i,t}^{\text{eff}}$ in \eqref{e1d}, we employ McCormick envelope constraints to reformulate it:
\begin{subequations}
\begin{align}
r_{i,t}^{\text{eff}} &\leq r_{i,t} \cdot \eta_{i}^{\text{max}}, \forall i \in \mathcal{I}, t \in \mathcal{T}  \label{e2a}\\
r_{i,t}^{\text{eff}} &\leq x_i^{max}R_{i}^{\text{max}} \cdot \eta_{i,t}^{\text{eff}}, \forall i \in \mathcal{I}, t \in \mathcal{T} \label{e2b}\\
r_{i,t}^{\text{eff}} &\geq r_{i,t} \cdot \eta_{i}^{\text{min}}, \forall i \in \mathcal{I}, t \in \mathcal{T} \label{e2c}
\end{align}
\end{subequations}

After the reformulation, the problem, $\mathcal{P}_1$, is transformed into a mixed-integer quadratic programming (MIQP) problem, $\mathcal{P}_2$:
\begin{subequations}
\begin{align}
&\mathop{\min}_{\mathbf{X}} \quad  \sum_{t \in \mathcal{T}} \{ P_{t}^{\text{imp}} \cdot \lambda_{t}^{\text{imp}} - P_{t}^{\text{exp}} \cdot \lambda_{t}^{\text{exp}}  \notag +\sum_{i \in \mathcal{I}} \delta_{i,t} \cdot C_{i}^{\text{penalty}} \\ 
&+\sum_{g \in \mathcal{G}} (\alpha_{g} \cdot p_{g,t} + \beta_{g} \cdot p_{g,t}^2 ) -\sum_{s \in \mathcal{S}}  P_{s,t}^{\text{dch}} \cdot \lambda_{t}^{\text{reg}} \notag \\
&+ \sum_{s \in \mathcal{S}} ( P_{s,t}^{\text{ch}} + P_{s,t}^{\text{dch}}) \cdot c_{s}^{\text{deg}}\} \cdot \Delta t   \\
&\mathbf{s.t.} \quad \eqref{e1b}-\eqref{e1c},\eqref{e1e}-\eqref{e1ac}, \eqref{e2a}-\eqref{e2c}, 
\end{align}
\end{subequations}

$\mathcal{P}_2$ is a centralized optimization problem, which requires the private data from all the data centers. This can raise privacy leakage concerns and meanwhile incur computational burdens. To mitigate these concerns, we propose the adaptive federated learning-to-optimization approach.

\section{Adaptive Federated Learning-to-Optimization Approach}
In this section, we present an adaptive federated learning approach that supports personalized and privacy-preserving learning and inference of data center decision variables. Utilizing locally inferred decisions, the original optimization problem can be reformulated into a simplified structure that necessitates only partial data sharing across data centers. To strengthen privacy guarantees while maintaining the integrity of shared variables, a double-aggregation mechanism is further proposed and analyzed.
\subsection{Adaptive Federated Learning Approach}
In the reformulated problem, $\mathcal{P}_2$, a data-driven approach via learning from historical optimal operating decisions conditioned on realized uncertainty can help inform the decision-making for hyperscale data centers. Historical decisions perform as the expert demonstrations, informing the learning process of the agent for each data center. For each data center, the environmental inputs are denoted as $u\in\mathbb{R}^{d_u}$, consisting of $\lambda_{t}^{\text{imp}}, \lambda_{t}^{\text{exp}}, \lambda_{t}^{\text{reg}}, q_{i,t}, u_{i,t}$. Under the environmental inputs, optimal decisions are denoted as $y\in\mathbb{R}^{d_y}$, consisting of $x_{i,t}, r_{i,t}, \eta_{i,t}^{\text{eff}}, T_{i,t}, r_{i,t}^{\text{eff}}, p_{i,t}^{\text{dyn}}, p_{i,t}^{\text{server}}, P_{i,t}^{\text{cool}}, Q_{i,t}^{\text{cool}},\delta_{i,t}$. To derive the mapping between environmental inputs and optimal decisions, a four-layer multilayer perceptron is employed:
\begin{subequations}
\begin{align}
&z_1 = d_1 u + b_1, \quad h_1 = \sigma(z_1), \quad \tilde h_1 = D_{p}\!\left(h_1\right),\\
&z_2 = d_2 \tilde h_1 + b_2, \quad h_2 = \sigma(z_2), \quad \tilde h_2 = D_{p}\!\left(h_2\right),\\
&z_3 = d_3 \tilde h_2 + b_3, \quad h_3 = \sigma(z_3), \quad \tilde h_3 = D_{p}\!\left(h_3\right),\\
& y = d_4 \tilde h_3 + b_4,
\end{align}
\end{subequations}

where $\sigma(\cdot)=\mathrm{ReLU}(\cdot)$ and $D_p$ denotes dropout with a given probability. For the adaptive federated learning, only parameters of the first few blocks, collected in the index set $\mathcal{K}$ (e.g., the first three blocks), are shared, i.e., $\{(d_1,b_1),(d_2,b_2),(d_3,b_3)\}$. Meanwhile, for each agent (data center), an ensemble of $N$ neural networks is designed to boost the generalization performance. For agent $i$ at the learning round $r$, parameters from all the $N$ neural networks in the ensemble model are averaged for the blocks in $\mathcal{K}$, respectively.
\begin{align}
\bar{d}^{(i)}_{r,\ell}
&= \frac{1}{N}\sum_{j=1}^{N} d^{(i,j)}_{r,\ell},
&
\bar{b}^{(i)}_{r,\ell}
&= \frac{1}{N}\sum_{j=1}^{N} b^{(i,j)}_{r,\ell}, \forall \ell \in \mathcal{K}.
\end{align}
Let $\theta_{r,i}=\{\bar{d}^{(i)}_{r,\ell}, \bar{b}^{(i)}_{r,\ell}\}_{\ell \in \mathcal{K}}$. The parameters, $\theta_{r,i}$, can be shared with the aggregator, i.e., the utility company. To further ensure the privacy of shared parameters, random weights are designed for each agent.
At the learning round $r$, suppose a total of $m$, $m=|\mathcal{I}_r|$, agents are involved in the federated learning. At this learning round, each agent $i$, $i \in \mathcal{I}_r$, generates its own secret random variable ${w}_{r,i}$ and splits the variable into $n$, $n \ge m$, shares, from ${w}_{r,i}^1$ to ${w}_{r,i}^n$ as:
\begin{subequations}
    \begin{align}
     {w}_{r,i}^1 = &{w}_{r,i} + \varphi_{i,1} Z_1+...+\varphi_{i,m-1} Z_1^{m-1} \label{sp1}\\
     {w}_{r,i}^2 = &{w}_{r,i} + \varphi_{i,1} Z_2+...+\varphi_{i,m-1} Z_2^{m-1} \\
        &  \vdots          \notag    \\
     {w}_{r,i}^n = &{w}_{r,i} + \varphi_{i,1} Z_n+...+\varphi_{i,m-1} Z_n^{m-1}, \label{sp2}
\end{align}
\end{subequations}
where $Z_1$, ..., $Z_{n}$ are positive integers, each corresponding to one agent. $\varphi_{i,1}$, ..., $\varphi_{i,m-1}$ are additional random variables generated by agent $i$. Each agent $i$ needs to send the split share ${w}_{r,i}^j$ to agent $j$, $j \in \mathcal{I}_r$ and $j \ne i$, through the secure communication channel, which can be realized via the CRT-Paillier \cite{liu2023privacy} or CKKS cryptosystems \cite{benaissa2021tenseal}. After receiving the split shares, each agent $j$ is able to compute the sum of split shares from all the $m$ agents by:
    \begin{align}
     &\Phi_j(Z_j)= \sum\limits_{i=1}^m({w}_{r,i}^j) \notag  \\
     &=\sum\limits_{i=1}^m({w}_{r,i})+\sum\limits_{i=1}^m(\varphi_{i,1})Z_j+...+\sum\limits_{i=1}^m(\varphi_{i,m-1}) Z_j^{m-1}.   \label{sp3} 
\end{align}
Then each agent $j$, $j \in \mathcal{I}_r$, sends the sum of split shares, $\Phi_j(Z_j)$, through the secure communication channel to the aggregator. Specifically, each agent encrypts the sum of split shares using the public key provided by the aggregator. When the aggregator receives the encrypted sum, its true value can be retrieved by decryption. The sum of random variables from all agents can be reconstructed by the aggregator using the Lagrangian interpolation method:
\begin{align}
     w_r=\sum\limits_{j=1}^m[\Phi_j(Z_j)\cdot\prod\limits_{h=1, h\neq j}^{m}\frac{Z_h}{Z_h-Z_j}]=\sum\limits_{i=1}^m({w}_{r,i}). \label{sp4}
\end{align}

 During the online phase, each agent $i$, $i \in \mathcal{I}_r$, masks its shared variables at the learning round $r$ as:
\begin{align}
    \hat{\theta}_{r,i}={w}_{r,i}\cdot\theta_{r,i} . \label{sp5}
\end{align}
Accordingly, the aggregator can employ the weighted parameters for the federated learning update:
\begin{align}
    \overline{\theta}_{r}=\frac{\sum\limits_{i \in \mathcal{I}_r}({\hat{\theta}_{r,i})}}{w_r}. \label{sp6}
\end{align}

After obtaining the updated parameters from the aggregators, the adaptive federated learning approach is proposed for each agent $i$, $i \in \mathcal{I}_r$, to decide whether the aggregate weights contribute to performance improvement for each agent based on the following criteria:
\begin{subequations}
\begin{align}
        &I_{r,i} = \text{clip}\left(\frac{\text{Mdn}(\delta_{r,i}^{\text{ind}}[-L:]) - \text{Mdn}(\delta_{r,i}^{\text{fed}}[-L:])}{\text{Mdn}(\delta_{r,i}^{\text{ind}}[-L:]) + \epsilon}, -1, 1\right) \label{sp7}\\        
        &T_{r,i} = \text{clip}\left(\text{Slp}(\delta_{r,i}^{\text{ind}}[-L:]) - \text{Slp}(\delta_{r,i}^{\text{fed}}[-L:]), -1, 1\right) \label{sp8}\\
        &S_{r,i} = \text{clip}\left(1 - \frac{\text{Std}(\delta_{r,i}^{\text{fed}}[-L:])}{\text{Std}(\delta_{r,i}^{\text{ind}}[-L:]) + \epsilon}, 0, 1\right) \label{sp9}\\
       &C_{r,i} = \kappa_1 \cdot \mathbbm{1}_{\{I_{r,i} > 0\}} + \kappa_2 \cdot \mathbbm{1}_{\{T_{r,i} > 0\}} + \kappa_3 \cdot S_{r,i}, \label{sp10}
\end{align}
\end{subequations}
where \text{Mdn}, \text{Slp}, and \text{Std} refer to the median, slope, and standard deviation of given data series. \eqref{sp7} refers to the relative reduction in the median error achieved by the federated learning approach compared to the independent learning, normalized by the independent median over the last $L$ rounds. $\delta_{r,i}^{\text{ind}}[-L:]$ and $\delta_{r,i}^{\text{fed}}[-L:]$ denote the composite loss values for independent and adaptive federated learning approaches, respectively. $I_{r,i} \ge 0$ means the federated approach performs better than the independent approach, and vice versa. \eqref{sp8} refers to the differences of linear slope between the two approaches over the last $L$ rounds. $T_{r,i} \ge 0$ means the federated approach’s slope is more negative and can improve the performance much faster than the independent approach. \eqref{sp9} refers to the volatility (standard deviation) of the federated approach compared to the independent one.  $S_{r,i}$ is larger when the federated learning is less volatile than the independent learning. $C_{r,i}$ is the weighted index. If the acceptance criteria are met, the weight parameters are updated via the momentum mixing manner.
                
If the acceptance criteria are not met, original local parameters, i.e., $\theta_{r,i}$, are not changed. Based on the updated model parameters for each agent, mean predictions and corresponding standard deviations from the ensemble model of each agent can be derived as:
\begin{subequations}
\begin{align}
y_{r,i}(u)&=\frac1N\sum_{n=1}^{N} f(u;\theta_{r,i}^{(n)}) \\
\sigma_{r,i}(u)&=\mathrm{Std}\big(\{f(u;\theta_{r,i}^{(n)})\}_{n=1}^{N}\big),
\end{align}
\end{subequations}
where $\theta_{r,i}^{(n)}$ refer to parameters of the $n$-th neural network for the agent $i$ at the learning round $r$. To account for the heterogeneity of the local dataset and to filter out potentially contaminated data from the aggregator, the acceptance rate is employed for each agent. If the acceptance rate is low for the agent $i$ after the first $k$ rounds, i.e., $\frac{T_{i,acc}}{\max(T_{i,acc} + T_{i,rej}, 1)} < 0.1$, where $T_{i,acc}$ and $T_{i,rej}$ denote the numbers of accepted and rejected aggregate updates, agent $i$ will opt out of federated learning for $k$ rounds. The momentum mixing strategy is proposed for the parameter updates:
\begin{subequations}
\begin{align}
&\alpha_{\text{base}} = 0.1 \cdot 0.95^{r}, \quad \alpha_{r,i} = \alpha_{\text{base}} \cdot C_{r,i} \label{sp11} \\
&v_{r,i} = \beta v_{r-1,i} + (1-\beta)(\theta_{r,i}-\bar\theta_{r})\label{sp12} \\
&\theta_{r+1,i} = \theta_{r,i} + \alpha_{r,i} v_{r,i}. \label{sp13}
\end{align}
\end{subequations}
The convergence guarantee of the proposed update scheme is further provided in Appendix.

\begin{figure}[!hbp]
	\centering
    \includegraphics[width=0.9\linewidth]{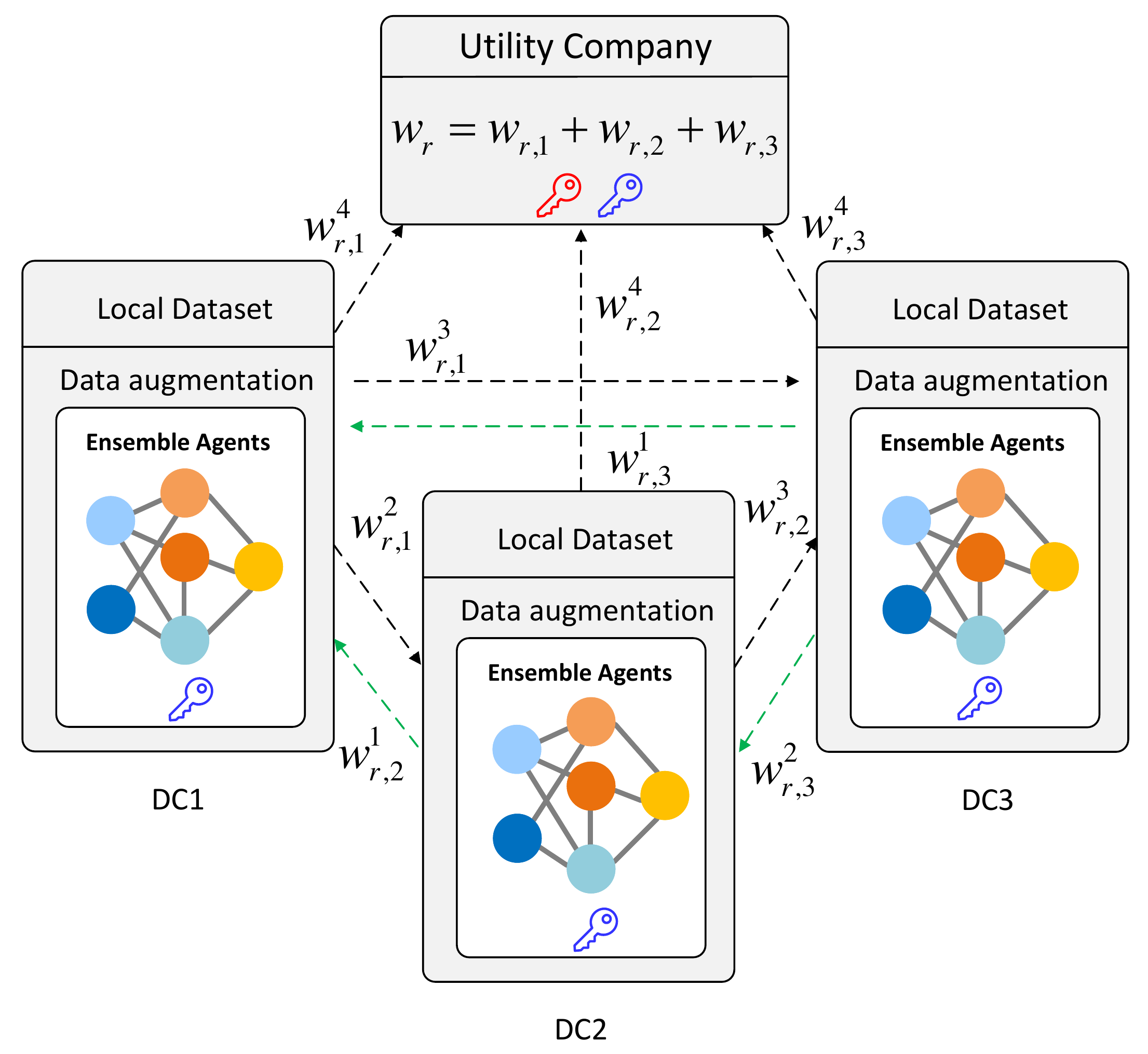}
	\caption{Adaptive federated learning (offline phase).}
	\label{fig2}
\end{figure}

\begin{figure}[!hbp]
	\centering
    \includegraphics[width=0.9\linewidth]{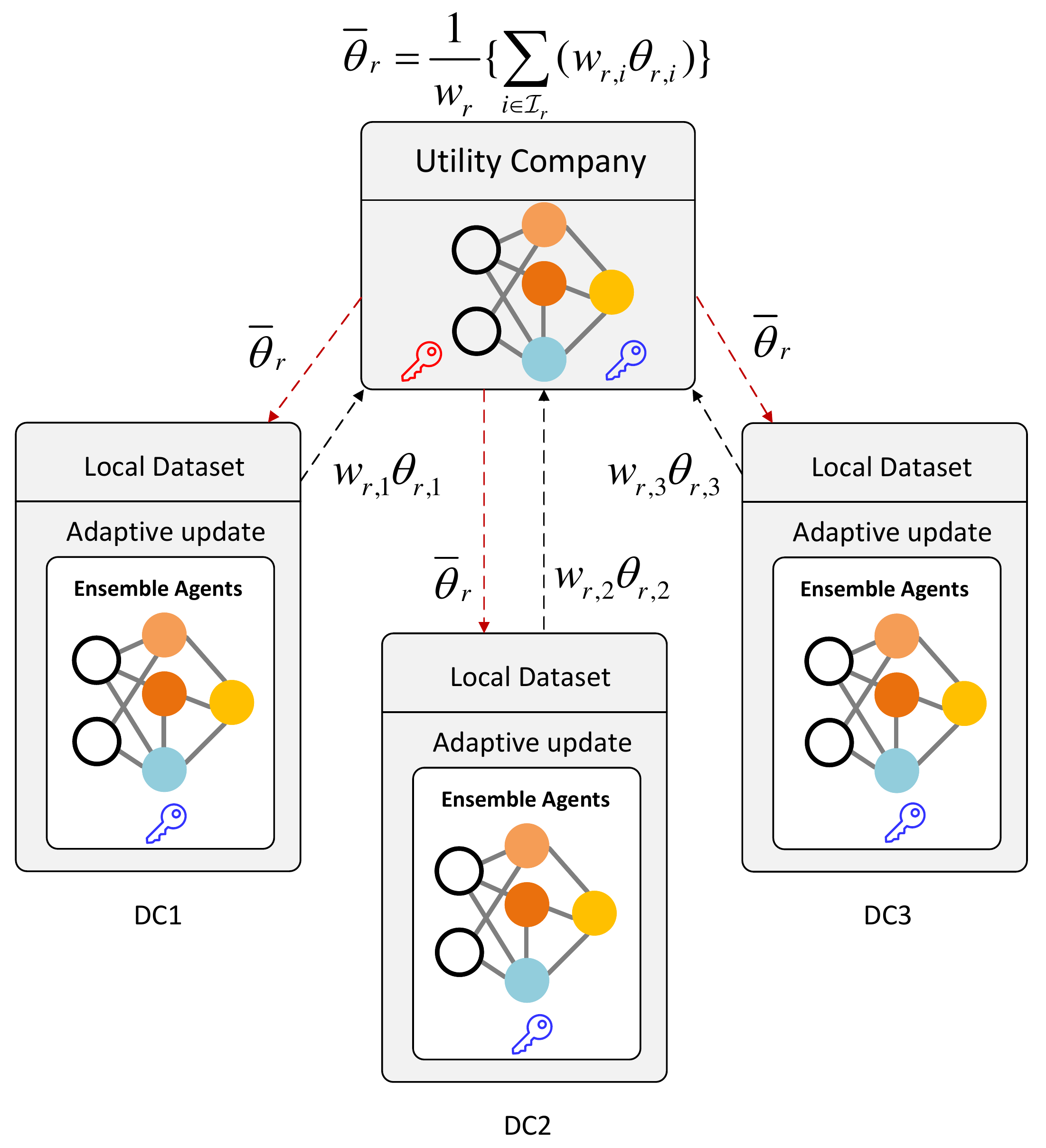}
	\caption{Adaptive federated learning (online phase).}
	\label{fig3}
\end{figure}

\begin{algorithm}[!t]
\caption{Adaptive Federated Learning}
\label{alg:enhanced-conditional-fl}
\begin{algorithmic}[1]
\STATE $\mathbf{Initialize}$ Agent $i$, $i \in \mathcal{I}$, with loss histories $\delta_i^{\text{ind}}$ and $\delta_i^{\text{fed}}$, thresholds $\tau_1 = 0.1, \tau_2 = 0.01, \tau_3 = 0.8, \kappa_1=0.4, \kappa_2=0.3$, and $\kappa_3=0.3$, window $L = 20$, momentum $\beta = 0.9$ and $v_i \gets \mathbf{0}$, $\epsilon = 10^{-6}$, $T_{i,acc} = 0$, $T_{i,rej} = 0$, and rounds $k = 5$
\STATE \textbf{Phase 0: Warmup Independent Training}
\FOR{each agent $i$, $i \in \mathcal{I}$, \textbf{in parallel}}
    \STATE Train for $T_{\text{warm}}$ steps with composite loss function:
    \STATE \quad $\delta_i^{\text{ind}} = 0.85 \cdot \text{MSE} + 0.15 \cdot \text{Huber} + \lambda \|\theta\|^2$
\ENDFOR
\STATE \textbf{Phase 1: Adaptive Federated Learning}
\FOR{round $r = 1$ to $R$}
    \STATE $\mathcal{I}_r \gets \{\text{agents are willing to participate}\}$
    
    \IF{$|\mathcal{I}_r| \geq 3$}        
        \STATE {\color{commentcolor}\textit{// Privacy-preserving aggregation}}
        \FOR{each agent $i$, $i \in \mathcal{I}_r$,}
            \STATE Generate and share $w_{r,i}$ following $\eqref{sp1}-\eqref{sp3}$
        \ENDFOR
        \STATE Reconstruct $w_{r}$  following $\eqref{sp4}$
        \STATE Calculate $\bar\theta_r \gets (\sum_{i \in \mathcal{I}_r} w_{r,i} \cdot \theta_{r,i})/w_{r}$
        
         \STATE {\color{commentcolor}\textit{// Warm-up federated training for participants}}
        \FOR{each agent $i$, $i \in \mathcal{I}_r$, \textbf{in parallel}}
            \STATE Train for $T_{\text{fed}}$ steps, append losses to $\delta_{r,i}^{\text{fed}}$
        \ENDFOR        
        \STATE {\color{commentcolor}\textit{// Conditional parameter update}}
        \FOR{each agent $i$, $i \in \mathcal{I}_r$,}
            \STATE \textbf{Compute adaptive acceptance criteria:}
            \IF{$(I_{r,i} > \tau_1 \textit{ or } T_{r,i} > \tau_2) \textit{ and } C_{r,i} > \tau_3$}
                \STATE {\color{commentcolor}\textit{// Accept update with momentum mixing}}
                \STATE \text{Update }$\theta_{r,i} \text{ following } \eqref{sp11}-\eqref{sp13}$
                \STATE $T_{i,acc} \gets T_{i,acc} + 1$
            \ELSE
                \STATE Continue with local parameters
                \STATE $T_{i,rej} \gets T_{i,rej} + 1$
            \ENDIF
            
            \STATE {\color{commentcolor}\textit{// Dynamic participation decision}}
            \IF{$r \geq k$ \AND $\frac{T_{i,acc}}{\max(T_{i,acc} + T_{i,rej}, 1)} < 0.1$}
                \STATE agent $i$ opts out of federation for next $k$ rounds
            \ENDIF
        \ENDFOR
    \ENDIF
    
    \STATE {\color{commentcolor}\textit{// Independent training for non-participants}}
    \FOR{each agent $i$, $i \in \mathcal{I} \setminus \mathcal{I}_r$}
        \STATE \quad Train for $T_{\text{fed}}$ steps, append losses to $\delta_{r,i}^{\text{ind}}$
    \ENDFOR
\ENDFOR
\STATE \textbf{return} Trained models for agents $\mathcal{I}$
\end{algorithmic}
\end{algorithm}
\vspace{-0.5cm}

\subsection{Federated Learning-to-Optimization Approach}
After the adaptive federated training, each agent/data center can perform local predictions for decision variables. Only decision variables for the energy storage systems, generators, and energy imports/exports remain to be derived. The optimization problem, $\mathcal{P}_2$, is approximately simplified to $\mathcal{P}_3$ as below:
\begin{subequations}
\begin{align}
&\mathop{\min}_{\mathbf{X}^{\dag}} \quad  \sum_{t \in \mathcal{T}} \{ P_{t}^{\text{imp}} \cdot \lambda_{t}^{\text{imp}} - P_{t}^{\text{exp}} \cdot \lambda_{t}^{\text{exp}}  \notag +\sum_{i \in \mathcal{I}} \delta_{i,t} \cdot C_{i}^{\text{penalty}} \\ 
&+\sum_{g \in \mathcal{G}} (\alpha_{g} \cdot p_{g,t} + \beta_{g} \cdot p_{g,t}^2 ) -\sum_{s \in \mathcal{S}}  P_{s,t}^{\text{dch}} \cdot \lambda_{t}^{\text{reg}} \notag \\
&+ \sum_{s \in \mathcal{S}} ( P_{s,t}^{\text{ch}} + P_{s,t}^{\text{dch}}) \cdot c_{s}^{\text{deg}}\} \cdot \Delta t  \\
&\mathbf{s.t.} \quad \eqref{e1l}-\eqref{e1n}, \eqref{e1v}-\eqref{e1ac}, 
\end{align}
\end{subequations}
where $\mathbf{X}^{\dag}$ consists of the decision variables including $z_{s,t}, \delta_{t}^{\text{imp}}, P_{t}^{\text{imp}}, P_{t}^{\text{exp}}, p_{g,t}, P_{s,t}^{\text{ch}}, P_{s,t}^{\text{dch}}$, and $E_{s,t}$. To facilitate the solution of $\mathcal{P}_3$, each agent is required to share its decision variables with the utility company, e.g., $p_{i,t}^{\text{dc}}$ and $\delta_{i,t}$. To preserve the privacy and integrity of the shared data, the verifiable secret data-sharing scheme with double aggregation scheme is proposed in the next section.

\subsection{Verifiable Private Data-Sharing Scheme}
For each agent $i$, the random variable $w_{r,i}$ is generated during the model training process and remains only known to that agent. To preserve privacy of the agent’s shared variable, e.g., $p_{i,t}^{dc}$, the agent adds a random mask $w_{r,i}$ before transmitting it to the utility, i.e., $p_{i,t}^{dc}+w_{r,i}$. In this manner, the privacy of the shared variable is well preserved. However, the shared data can be contaminated by false data injections from adversaries, which raises data integrity concerns. It remains challenging to detect the false data injections when the shared variables are kept secret. To further deal with this dilemma, the double aggregation mechanism, where two sets of masked/encrypted data are shared with the aggregator for the cross verification, is proposed to simultaneously ensure the privacy and integrity of shared variables. Without loss of generality, we assume that all the data centers participate in the data-sharing stage, i.e., $\mathcal{I}_r=\mathcal{I}$. To realize this, the utility company needs to broadcast its public key, i.e., $PK_u$, to all the agents. At the $S_1$ step, the utility company sends the encrypted large random coefficient $\pi$, $\pi \ge 10$, using its public key $PK_u$. At the $S_2$ step, each agent will perform the encrypted multiplication for $\pi$ and $p_{i,t}^{dc}+w_{r,i}$, where $w_{r,i}$ is the random variable for agent $i$. Then the encrypted addition is performed with an additional variable $\gamma_{i,t}$. The encrypted result, $\gamma_{i,t}+\pi(p_{i,t}^{dc}+w_{r,i})$, along with the masked data $p_{i,t}^{dc}+w_{r,i}$ and $\gamma_{i,t}+w_{r,i}$, is sent to the utility company. At the $S_3$ step, the utility company decrypts the received result, i.e., $\gamma_{i,t}+\pi(p_{i,t}^{dc}+w_{r,i})$, using its private key $SK_u$. Then $\Omega$ can be derived through the aggregation of decrypted results and subtraction of $\pi w_r$. Meanwhile, $\sum_{i \in \mathcal{I}} p_{i,t}^{dc}$ can be derived by the aggregation of $p_{i,t}^{dc}+w_{r,i}$ and the subtraction of $w_r$. $\gamma=\sum_{i \in \mathcal{I}} \gamma_{i,t}$ can be derived via the same manner. At the $S_4$ step, the utility company broadcasts a set of variables $\Omega, \gamma, \pi\sum_{i \in \mathcal{I}} p_{i,t}^{dc}$ to all the agents. Each agent can perform the verification via $\Omega ?= \gamma + \pi\sum_{i \in \mathcal{I}} p_{i,t}^{dc}$. To tolerate errors from the encryption and decryption, a relaxed condition can be applied, i.e., $abs(\Omega-\gamma - \pi\sum_{i \in \mathcal{I}} p_{i,t}^{dc}) ? < \psi$. The tolerance $\psi$ is calibrated above the numerical noise floor of the encrypted computations, as detailed in Section IV-D. The overall online workflow, from local model inference to the verifiable solution of $\mathcal{P}_3$, is summarized in Algorithm \ref{alg:verifiable}.

\begin{algorithm}[!t]
\caption{Verifiable Federated Learning-to-Optimization (Online Phase)}
\label{alg:verifiable}
\begin{algorithmic}[1]
\STATE $\mathbf{Initialize}$ trained ensemble parameters $\theta_{r,i}$ for each agent $i$, $i \in \mathcal{I}$; utility key pair $(PK_u, SK_u)$; public points $Z_1, ..., Z_n$; tolerance $\psi$
\STATE {\color{commentcolor}\textit{// Process 1: Model inference}}
\FOR{each agent $i$, $i \in \mathcal{I}$, \textbf{in parallel}}
    \STATE Predict local decision variables, e.g., $p_{i,t}^{\text{dc}}$ and $\delta_{i,t}$, $\forall t \in \mathcal{T}$, via the ensemble model
\ENDFOR
\STATE {\color{commentcolor}\textit{// Process 2: Mask generation and secret sharing}}
\STATE Each agent $i$ generates $w_{r,i}$ and distributes the split shares following \eqref{sp1}-\eqref{sp3}
\STATE The utility company reconstructs $w_r$ following \eqref{sp4}
\STATE {\color{commentcolor}\textit{// Process 3: Encrypted computation and verification}}
\STATE $S_1$: the utility company broadcasts $Enc(PK_u, \pi)$, $\pi \ge 10$
\FOR{each agent $i$, $i \in \mathcal{I}$, \textbf{in parallel}}
    \STATE $S_2$: send $p_{i,t}^{\text{dc}}+w_{r,i}$, $\gamma_{i,t}+w_{r,i}$, and $Enc(PK_u, \gamma_{i,t}+\pi(p_{i,t}^{\text{dc}}+w_{r,i}))$, $\forall t \in \mathcal{T}$
\ENDFOR
\STATE $S_3$: the utility company decrypts and aggregates the received data to derive $\Omega$, $\gamma$, and $\sum_{i \in \mathcal{I}} p_{i,t}^{\text{dc}}$
\STATE $S_4$: the utility company broadcasts $\{\Omega, \gamma, \pi\sum_{i \in \mathcal{I}} p_{i,t}^{\text{dc}}\}$; each agent verifies $abs(\Omega-\gamma - \pi\sum_{i \in \mathcal{I}} p_{i,t}^{dc}) < \psi$
\IF{the verification fails}
    \STATE Discard the shared data and trigger countermeasures
\ELSE
    \STATE {\color{commentcolor}\textit{// Process 4: Simplified optimization}}
    \STATE The utility company solves $\mathcal{P}_3$ via branch and bound to derive the dispatch decisions $\mathbf{X}^{\dag}$
\ENDIF
\STATE \textbf{return} Dispatch decisions $\mathbf{X}^{\dag}$ and locally predicted decision variables
\end{algorithmic}
\end{algorithm}

\begin{figure}[!hbp]
	\centering
    \includegraphics[width=0.9\linewidth]{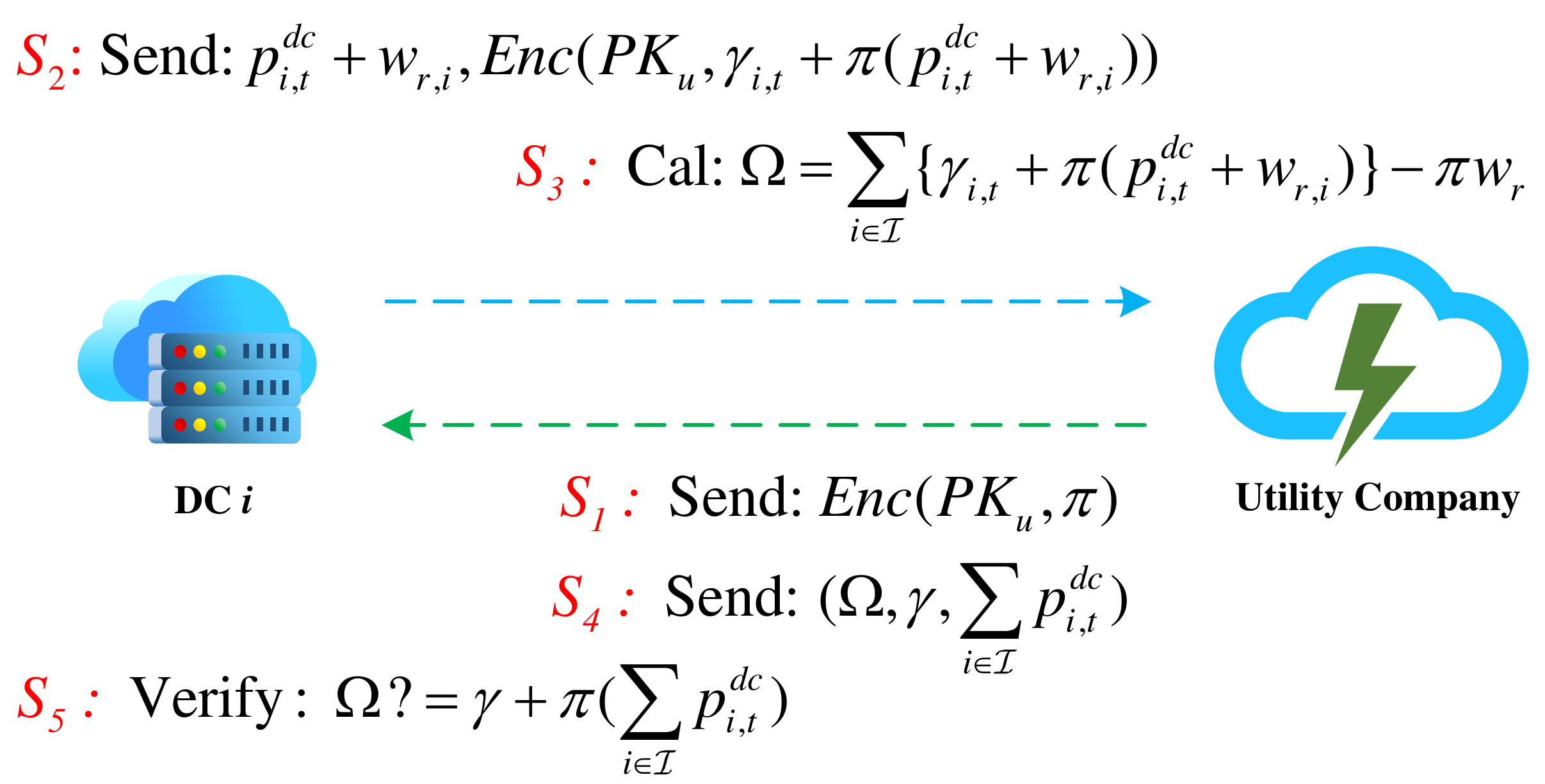}
	\caption{Verifiable data sharing via double aggregation mechanism.}
	\label{fig4}
	\vspace{-0.5cm}
\end{figure}

\subsection{Privacy and Security Analysis}
We first formalize the threat model under which the proposed approach operates, and then analyze its security guarantees.

{\bf Threat Model:} Two types of entities participate in the proposed framework: the agents (data centers), $i \in \mathcal{I}$, and the aggregator, i.e., the utility company. Both types of entities are modeled as honest-but-curious: they follow the designed protocols faithfully, yet may attempt to infer the private information of other participants, e.g., local datasets, model parameters $\theta_{r,i}$, secret masks $w_{r,i}$, and decision variables $p_{i,t}^{\text{dc}}$, from the messages they legitimately receive. Up to $m-1$ agents may collude by pooling their received information, whereas the aggregator is assumed not to collude with any agent. Since each colluding agent holds only one share of the honest agent $i$'s degree-$(m-1)$ polynomial in \eqref{sp1}-\eqref{sp2}, any coalition of at most $m-1$ agents faces an underdetermined system and cannot recover the secret mask $w_{r,i}$. Beyond the curious internal entities, we further consider external adversaries that launch false data injection attacks. They can tamper with the masked or encrypted variables transmitted between agents and the aggregator, i.e., $p_{i,t}^{\text{dc}}+w_{r,i}$, $\gamma_{i,t}+w_{r,i}$, and $Enc(PK_u, \gamma_{i,t}+\pi(p_{i,t}^{\text{dc}}+w_{r,i}))$, including coordinated joint injections across multiple messages from the same agent. During the learning stage, the aggregate model updates returned by the aggregator may likewise be contaminated. All adversaries are assumed to be computationally bounded. They cannot break the semantic security of the adopted CRT-Paillier and CKKS cryptosystems, cannot access the aggregator's private key $SK_u$, and hence cannot recover the verification coefficient $\pi$ from its ciphertext $Enc(PK_u,\pi)$. In addition, the secret mask $w_{r,i}$ is freshly generated by agent $i$ at each learning round, at least three agents participate in each aggregation round, i.e., $|\mathcal{I}_r| \ge 3$, and the aggregator only learns the intended aggregate quantities, e.g., $w_r$, $\overline{\theta}_{r}$, and $\sum_{i \in \mathcal{I}} p_{i,t}^{\text{dc}}$, rather than any individual value.

Under this threat model, the proposed approach pursues three security goals: i) \textit{privacy}, i.e., individual secret masks, model parameters, and shared decision variables shall not be disclosed to the aggregator, other agents, or external adversaries; ii) \textit{integrity}, i.e., any false data injection that perturbs the shared variables by no less than the tolerance $\psi$ shall be detected through the double aggregation verification; and iii) \textit{robustness}, i.e., contaminated aggregate model updates shall be filtered out by the acceptance criterion in \eqref{sp7}-\eqref{sp10}. Denial-of-service attacks and the poisoning of agents' local raw datasets are beyond the scope of this work. The integrity guarantee is formalized as follows.

{\bf Proposition 1:} The proposed verifiable approach can defend against joint false data injection attacks from adversaries.

\textit{Proof:} Suppose one agent $i$ is being attacked by adversaries, and $\varrho, \varrho \ge \psi$, is injected to the shared variable $p_{i,t}^{dc}+w_{r,i}$. Then three variables are sent to the utility company, i.e., $p_{i,t}^{dc}+w_{r,i}+\varrho$, $\gamma_{i,t}+w_{r,i}$, $Enc(PK_u, \gamma_{i,t}+\pi(p_{i,t}^{dc}+w_{r,i}))$. After the aggregation, the final verification step would be $abs(\Omega-\gamma - \pi\sum_{i \in \mathcal{I}} p_{i,t}^{dc}-\pi \varrho) > \psi$. This will not pass the verification. For the joint attacks, false data are simultaneously injected to multiple shared variables for agent $i$. The contaminated data can become $p_{i,t}^{dc}+w_{r,i}+\varrho$ and $\gamma_{i,t}+w_{r,i}-\varrho$. Encrypted ciphertext of $\gamma_{i,t}+\pi(p_{i,t}^{dc}+w_{r,i})$ is sent to the utility company, ensuring the integrity of $\Omega$. Meanwhile, $\pi$ is only known to the utility company and only its ciphertext, $Enc(PK_u,\pi)$, is shared with agents. This constrains the adversary's ability to design more sophisticated false data injection attacks by leveraging the information from $\pi$. The final verification step would be $abs(\Omega-\gamma + \varrho - \pi\sum_{i \in \mathcal{I}} p_{i,t}^{dc}-\pi \varrho) > \psi$. This will not pass the verification. On the other hand, if the encrypted ciphertext from agent $i$ is contaminated, the verification step will also not be passed due to the confidentiality of the utility company’s private key. Through the designed verifiable secret sharing scheme, the privacy and integrity of the shared variables are simultaneously guaranteed. To mitigate negative impacts from the detected false data injection attacks, countermeasures can be applied, including the online tensor mitigation approach \cite{liu2025byzantine} and Blockchain technology \cite{singh2024blockchain}.

\subsection{Computational Complexity Analysis}
To characterize how the computational burden grows with the process-level input sizes, we analyze the step-wise time complexity of the four processes in Algorithm \ref{alg:verifiable}. Processes 1), 3), and 4) correspond to the model inference, encryption and verification, and learning-to-optimization timing measurements reported in Fig. \ref{fig8}, respectively, while process 2) is not separately profiled since it involves only $O(m^2)$ scalar operations.

\textit{1) Model inference:} each agent evaluates an ensemble of $N$ four-layer MLPs over $|\mathcal{T}|$ periods. One forward pass costs $\sum_{\ell=1}^{4} H_{\ell-1}H_{\ell}$ multiply--accumulate operations, where $H_0=d_u$, $H_4=d_y$, and $H_1, H_2, H_3$ denote the hidden widths. The inference complexity is therefore $O(N|\mathcal{T}|\sum_{\ell}H_{\ell-1}H_{\ell})$ per agent, which is linear in the ensemble size and the horizon, and independent of the number of agents since all inferences run locally in parallel.

\textit{2) Mask generation and secret sharing:} evaluating the $n$, $n \ge m$, shares of the degree-$(m-1)$ polynomial in \eqref{sp1}-\eqref{sp2} costs $O(nm)$ operations per agent using Horner's rule, the pairwise share distribution incurs $O(m^2)$ messages in total, and the Lagrangian interpolation in \eqref{sp4} costs $O(m^2)$ operations at the aggregator. This stage thus scales quadratically with the number of participating agents, yet it is independent of the model and data sizes because only scalar masks are shared.

\textit{3) Encrypted computation and verification:} for the CKKS scheme with polynomial modulus degree $N_p$, one encryption or decryption costs $O(N_p \log N_p)$ operations, and so does one ciphertext--plaintext multiplication or addition. Since each ciphertext packs up to $N_p/2$ values, all $|\mathcal{I}||\mathcal{T}|$ shared variables occupy only $\lceil 2|\mathcal{I}||\mathcal{T}|/N_p \rceil$ ciphertexts. At the $S_2$ step, each agent performs $O(\lceil 2|\mathcal{T}|/N_p \rceil N_p \log N_p)$ operations; at the $S_3$ step, the utility company performs $O(m)$ ciphertext additions and one decryption; and the plaintext verification at the $S_4$ step costs $O(|\mathcal{T}|)$ per agent. The cryptographic overhead therefore grows linearly with both $m$ and $|\mathcal{T}|$.

\textit{4) Optimization:} the simplified problem $\mathcal{P}_3$ contains $(1+|\mathcal{S}|)|\mathcal{T}|$ binary variables and $O((|\mathcal{G}|+|\mathcal{S}|)|\mathcal{T}|)$ continuous variables, so the branch-and-bound method explores at most $O(2^{(1+|\mathcal{S}|)|\mathcal{T}|})$ nodes in the worst case, each solving a convex quadratic programming relaxation of polynomial complexity. In contrast, the original problem $\mathcal{P}_2$ additionally involves $|\mathcal{I}||\mathcal{T}|$ bounded integer variables $x_{i,t} \in \{0,...,x_i^{max}\}$ together with $O(|\mathcal{I}||\mathcal{T}|)$ extra continuous variables and constraints, which inflates the worst-case search space by a factor of up to $\prod_{i \in \mathcal{I}}(x_i^{max}+1)^{|\mathcal{T}|}$. Eliminating these integer variables through local inference is the dominant source of the acceleration observed in Fig. \ref{fig8}, whereas the privacy-preserving layers only add the linear and quadratic overheads of processes 2) and 3). Consequently, the proposed approach scales gracefully as the numbers of data centers and operation periods increase, with the optimization stage remaining the only combinatorial bottleneck.

\section{Numerical Simulations}
\subsection{Simulation Setup}
The code is implemented in Python 3.9 and executed on a computing platform equipped with 15 vCPUs (Intel® Xeon® Platinum 8474C) and an NVIDIA RTX 4090D GPU with 24 GB memory. The commercial solver, Gurobi V11.0 \cite{gurobi}, is employed to solve the optimization problems. The datasets used in this study, including market signals, load profiles, and operational data for the data centers, storage systems, and generators, are accessible at \cite{junhong2025}. For the federated learning, a four-layer MLP with hidden widths of [256, 1024, 256], ReLU activations, and a dropout rate of 0.15 after each hidden layer are used. Training is performed in two stages: an independent warm-up phase (1500 local steps with the batch size of 512) followed by conditional federated training using momentum mixing aggregation (25 global rounds, 800 local steps per round). Model updates were conditionally accepted based on validation performance improvements. The fully homomorphic encryption scheme (CKKS) from the tenseal library \cite{benaissa2021tenseal} is employed for encrypted computation. The encryption context is initialized with a polynomial modulus degree of 8192, providing approximately 128-bit security, and a coefficient modulus chain of [60, 40, 40, 60] bits, allowing up to two multiplications. The global scaling factor is set to $2^{40}$ to maintain numerical precision.

\subsection{Accuracy and Optimality}
\begin{figure}[!hbpt]
	\centering
    \includegraphics[width=0.9\linewidth]{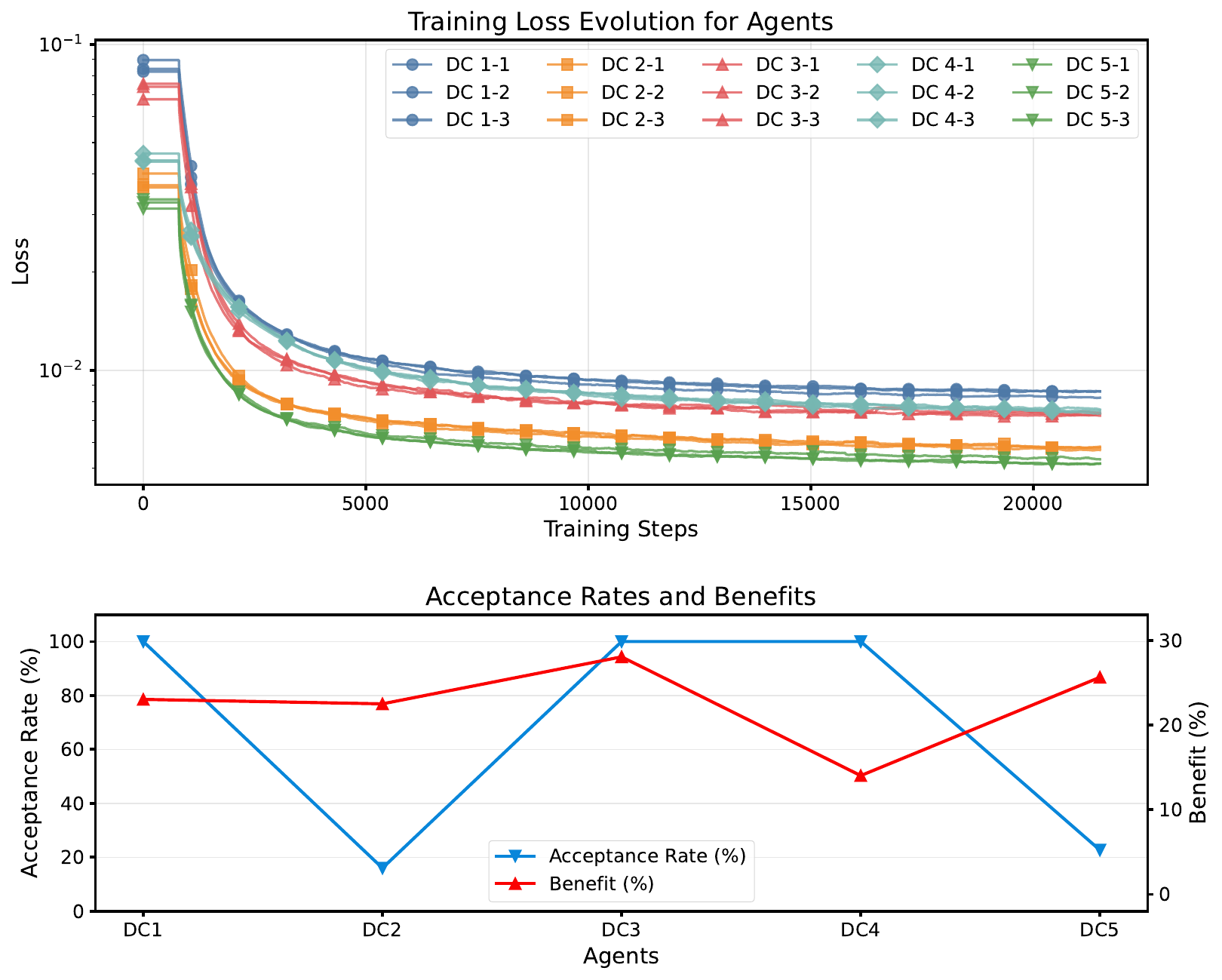}
	\caption{Training results for the adaptive federated learning.}
	\label{fig5}
	\vspace{-0.5cm}
\end{figure}

The accuracy of the proposed adaptive federated learning is validated by comparing its performance against two benchmark approaches: independent learning and default federated learning. As shown in Fig. \ref{fig5}, the normalized training loss decreases progressively and stabilizes after approximately 20000 training steps. The acceptance rates of data centers 1, 3, and 4 are observed to be close to 100\%, indicating nearly full acceptance of aggregate model updates. Data centers 2 and 5 maintain an acceptance rate of approximately 20\%, which indicates the heterogeneity of the datasets from different data centers. The normalized root mean squared error (NRMSE) is employed to quantify the scaled typical magnitude of prediction errors, as defined in equation \eqref{eq_r0}, where $n$ is the number of test samples, and $y_{i,k}$ and $\hat{y}_{i,k}$ denote the true and predicted values of the $k$-th decision variable for sample $i$, with $\bar{y}$ as the mean of the true values. The benefit of adaptive federated learning is computed based on the relative change in NRMSE compared with independent learning, as defined in equation \eqref{eq_r1}. The overall benefit improvement across all data centers ranges between 15\% and 30\% compared to the independent learning approach, as given in Fig. \ref{fig5}. The coefficient of determination, i.e., $R^2$, is further employed to measure the relative performance gain over a baseline model that predicts the mean of decision variables, as defined in equation \eqref{eq_r2}. On the test dataset, the proposed adaptive federated learning approach achieves the best prediction performance compared with independent learning and default federated learning. As given in Table \ref{tab1}, the proposed adaptive federated learning consistently maintains a coefficient of determination, $R^2$, higher than $0.95$, which is assessed through all the decision variables $y\in\mathbb{R}^{d_y}$. Under cyberattack scenarios, each data center could reject the contaminated model parameters received from the utility company, operate independently, and still maintain a lower-bounded level of performance.

\vspace{-0.2cm}

\begin{subequations}
\begin{align}
&NRMSE \;=\; \frac{1}{d_y} \sum_{k=1}^{d_y} \frac{\sqrt{\frac{1}{n}\sum_{i=1}^{n}(y_{i,k}-\hat y_{i,k})^2}}{\text{Std}(y_{k})+10^{-8}}, \label{eq_r0} \\
&Benefit(\%) \;=\; 100 \cdot \frac{NRMSE_{ind}-NRMSE_{fed}}{NRMSE_{ind}}, \label{eq_r1}\\
&R^2 \;=\; 1 - \frac{\sum_{i=1}^n (y_i-\hat y_i)^2}{\sum_{i=1}^n (y_i-\bar y)^2}. \label{eq_r2}
\end{align}
\end{subequations}

\vspace{-0.2cm}

\begin{table}[htbp]
	\centering
	\caption{Performance of Different Methods ($R^2$)}
	\label{tab1}
	\begin{tabular}{cccccc}
		\toprule  
		&DC1 &DC2 &DC3 & DC4 & DC5 \\ 
		\cmidrule(r){2-6}
		{Independent Learning}&0.920 &0.630&0.959 &0.941 &0.916\\	
        {Default FL}&0.591&0.807&0.797 &0.642&0.935\\
        {Adaptive FL}&\textbf{0.951}&\textbf{0.988}&\textbf{0.988} &\textbf{0.984}&\textbf{0.988}\\ 
		\bottomrule 
	\end{tabular}
\end{table}

\begin{figure}[!hbpt]
	\centering
    \includegraphics[width=0.9\linewidth]{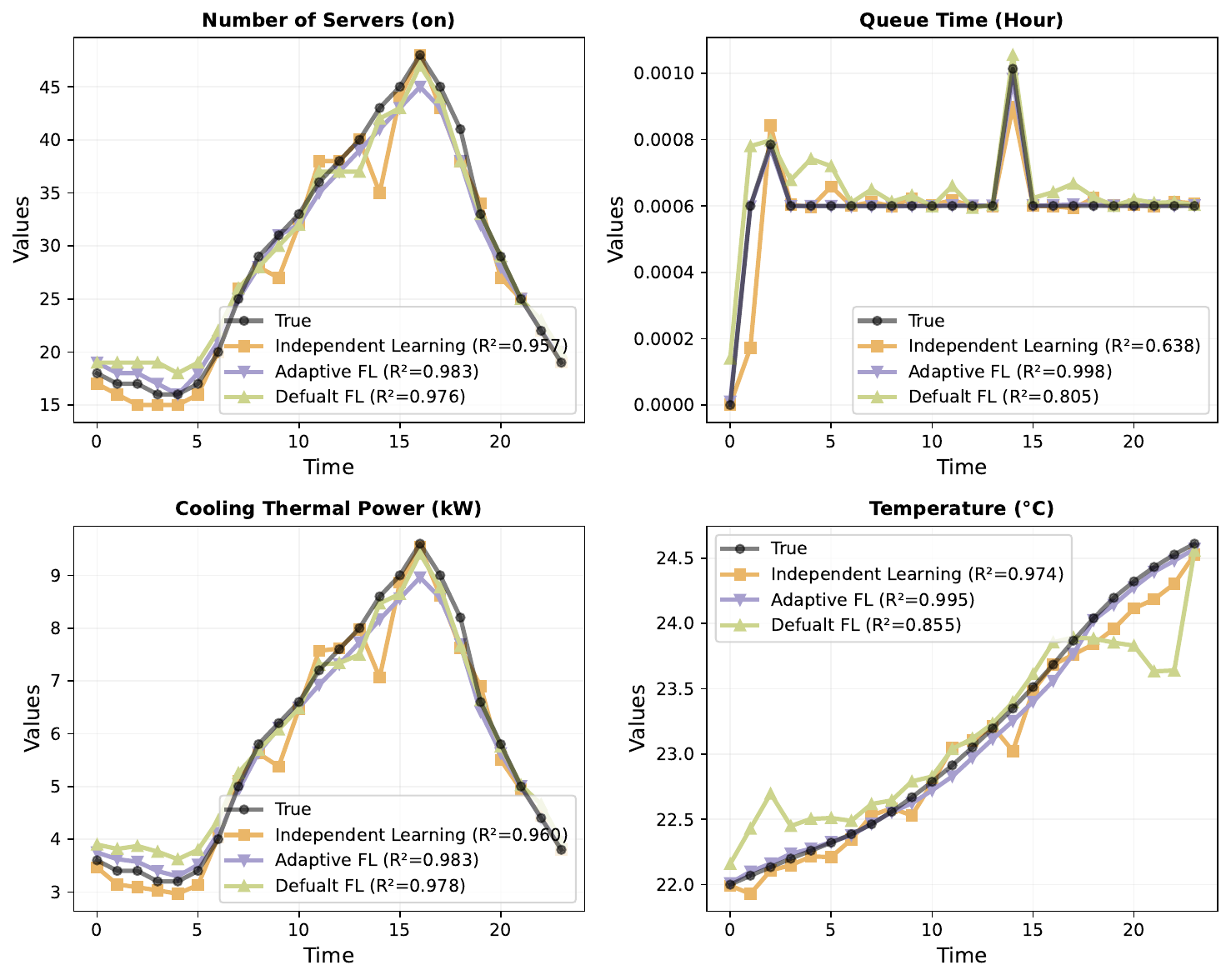}
	\caption{Prediction performance for data center 5.}
	\label{fig6}
\end{figure}

\begin{figure}[!hbpt]
	\centering
    \includegraphics[width=0.9\linewidth]{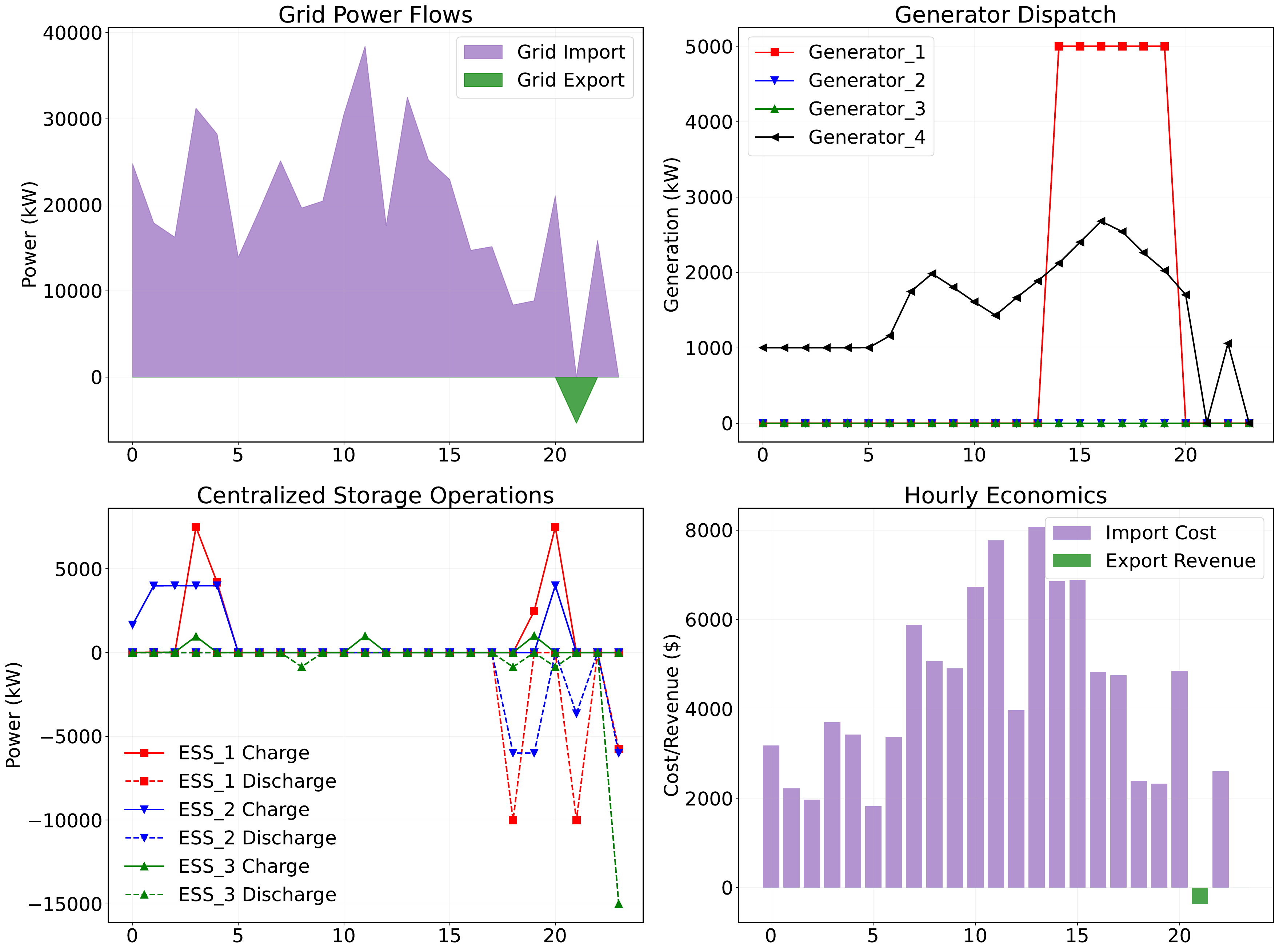}
	\caption{System states from the adaptive federated learning-to-optimization.}
	\label{fig7}
	\vspace{-0.5cm}
\end{figure}

The trained model at each data center enables the independent derivation of system states, including the committed number of servers, queue time, cooling thermal power, temperature, etc., as illustrated in Fig. \ref{fig6}. The adaptive federated learning approach produces system states that align more closely with the true values, resulting in higher $R^2$ scores, which is assessed independently for each decision variable. After deriving the states for each data center, the partial decision variables including $p_{i,t}^{\text{dc}}$ and $\delta_{i,t}$ can be sent to the utility company via the proposed secure and privacy-preserving double aggregation approach. To ensure the shared variables between data centers and the utility company are not contaminated by adversaries, the utility company is responsible for broadcasting the required information back to each data center for verification. Once the verification is passed, the simplified optimization problem can be solved to determine the optimal decision variables associated with the energy storage systems, generators, and grid imports and exports. As in Fig. \ref{fig7}, the system needs to import additional electricity from the wholesale market during all hours except the 21st hour. During the 18th to 21st hours, the battery storage units discharge energy to supply ancillary services and support system operations. The component costs derived from  different approaches are summarized in Table \ref{tab2}, where M0 denotes the reference centralized optimization, M1 represents the independent learning-assisted optimization, M2 is the default federated learning-assisted optimization, and M3 corresponds to the proposed adaptive federated learning-assisted optimization. Error$_{1-3}$ correspond to the relative errors obtained from methods M$_{1-3}$, respectively, evaluated with respect to results from method M$_0$. The component costs in Table \ref{tab2} refer to the energy import and export cost, generation cost, service delay penalty cost, ancillary service costs, charging and discharge costs for the battery, correspondingly. From the results, the proposed M$_3$ almost achieves the lowest relative errors for all component costs as denoted by Error$_3$. The relative errors from the aggregate costs for all the three methods are 0.12\%, 0.07\%, and 0.08\%, respectively. For individual components, M$_2$ exhibits both larger positive and negative deviations compared with the other methods, which contribute to a lower overall relative error. In contrast, the proposed approach, i.e., M$_3$, achieves balanced component-wise and overall errors, demonstrating its competitively near-optimal accuracy in the system optimization.

\begin{table}[htbp]
	\centering
	\caption{Component Costs from the Federated Learning-to-Optimization Approach (Dollar)}
	\label{tab2}
	\scriptsize
	\begin{tabular}{ccccccc}
		\toprule  
		&$Cost_1$ &$Cost_2$ &$Cost_3$ &$Cost_4$ & $Cost_5$ & All\\ 
		\cmidrule(r){2-7}
            {M0}&97292.73 &13236.12&1.84 &-46689.00  &6549.95 & 70391.63\\	
		    {M1}&97197.80 &13241.69&3.80 &-46688.95  &6550.15 & 70304.49\\	
            {M2}&97224.50&13250.09&2.80&-46688.76  &6555.38 & 70344.01\\
            {M3}&{97230.57}&{13241.69}&{0.93} &{-46688.95}  &{6551.53} & 70335.77\\ 
            {$\text{Error}_{1}$}&0.10\% &0.04\%&106.52\% &0.00\% &\textbf{0.00\%} & 0.12\%\\	
            {$\text{Error}_{2}$}&0.07\%&0.11\%&52.17\% &0.00\% &0.08\% &\textbf{0.07\%}\\
            {$\text{Error}_{3}$}&\textbf{0.06\%}&\textbf{0.04\%}&\textbf{49.46\%} &\textbf{0.00\%}&0.02\% & 0.08\%\\ 
		\bottomrule 
	\end{tabular}
	\vspace{-0.5cm}
\end{table}

\subsection{Computational Efficiency}
\begin{figure}[!hbpt]
	\centering
    \includegraphics[width=0.89\linewidth]{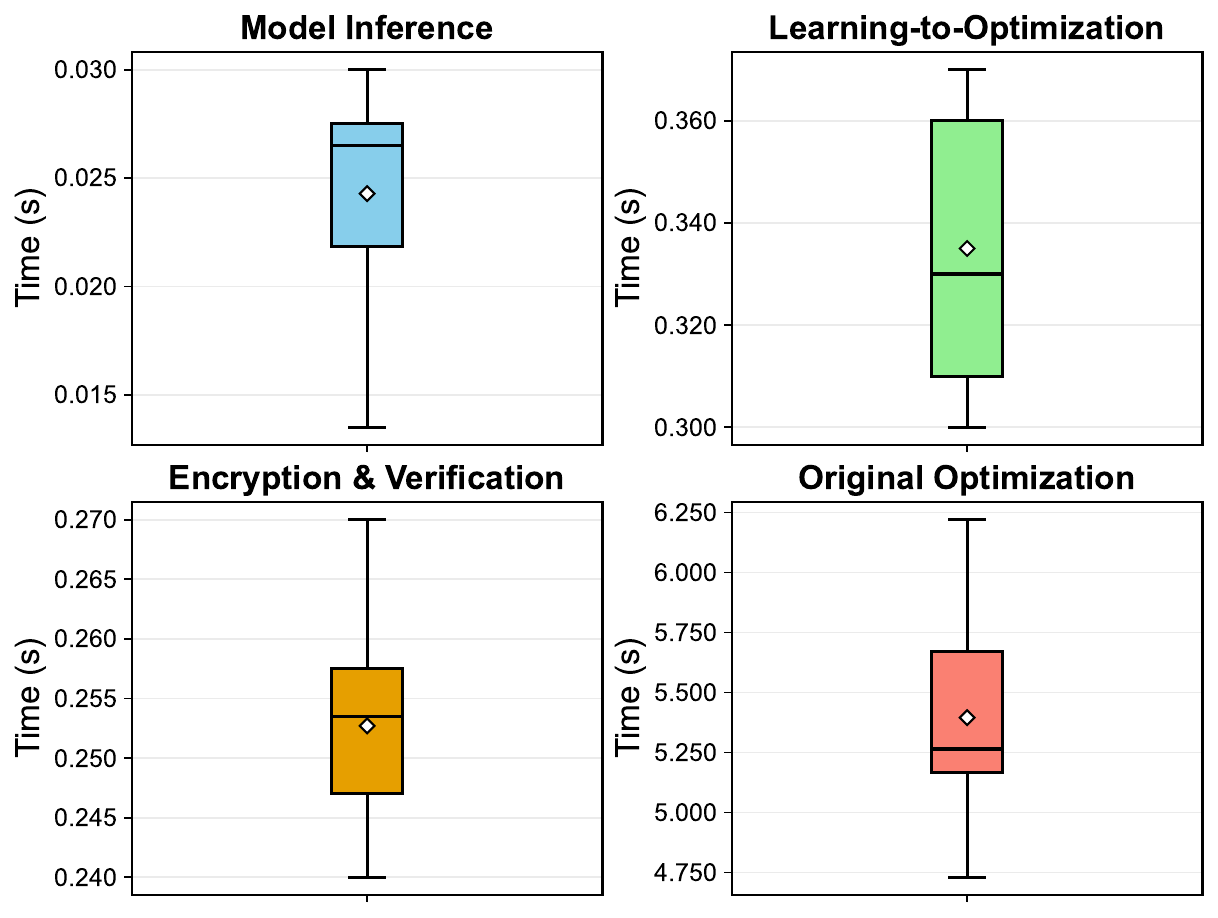}
	\caption{Time consumption for different approaches.}
	\label{fig8}
	\vspace{-0.5cm}
\end{figure}

The computational efficiency of the proposed approach is further validated. Ten simulation instances are conducted for both the proposed approach and the original centralized optimization approach. As shown in Fig. \ref{fig8}, the model inference for predicting data center decision variables takes approximately 0.025~s. Meanwhile, the federated learning-to-optimization takes approximately 0.330~s. The overall encrypted computation and verification step consume an additional 0.255~s. Overall, the proposed approach requires about 0.61~s per run. Comparably, the original centralized optimization approach consumes about 5.30~s to complete one run, which is about 8.69 times that of the proposed approach. These results demonstrate that the proposed approach is computationally lightweight and efficient, making it effective in reducing computation time for large-scale data center optimization.

\subsection{Adversarial Evaluation}
To validate the security guarantees under realistic attacks, we conduct two adversarial evaluations aligned with the threat model: false data injection against the verifiable data-sharing scheme in the optimization stage, targeting the integrity goal; and model poisoning against the adaptive federated learning, targeting the robustness goal.

\begin{figure*}[!hbpt]
	\centering
	\includegraphics[width=0.96\linewidth]{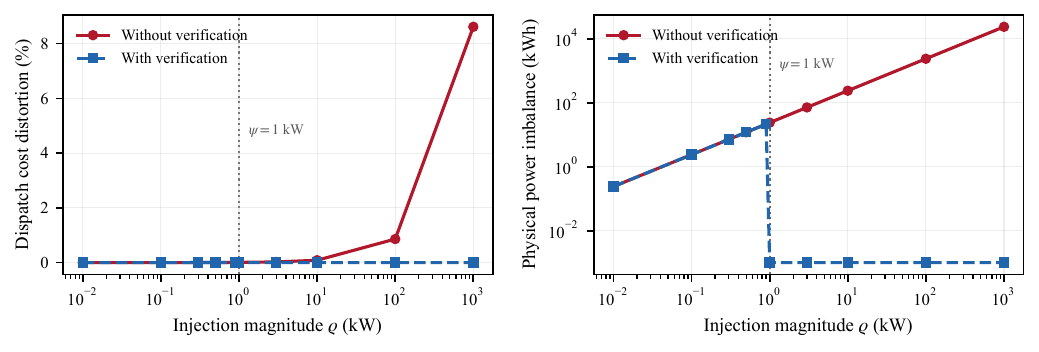}
	\caption{Impact of false data injection on the optimization stage versus injection magnitude $\varrho$: dispatch cost distortion (left) and physical power imbalance (right), with and without the verifiable double aggregation ($\psi=1$~kW).}
	\label{fig:poison_opt}
\end{figure*}

\subsubsection{Integrity under false data injection}
The verifiable double aggregation scheme (Algorithm \ref{alg:verifiable}) is implemented with the CKKS parameters of Section IV-A. The shared operational values $p_{i,t}^{\text{dc}}$ range from $8.7$ to $399.9$~kW across the five data centers. Over 200 attack-free executions of the $S_1$--$S_4$ pipeline, the peak verification residual $abs(\Omega-\gamma - \pi\sum_{i \in \mathcal{I}} p_{i,t}^{dc})$ per run has a maximum of $1.29\times10^{-2}$ and a mean of $7.08\times10^{-3}$, reflecting the intrinsic numerical error of the encrypted computations. The tolerance is therefore calibrated to $\psi=1$, roughly two orders of magnitude above this noise floor, which yields zero false positives over the 200 attack-free runs while remaining far below the magnitude of the shared values. Each pipeline execution takes $0.272\pm0.015$~s, consistent with the encryption and verification timing reported in Fig. \ref{fig8}.

We then inject false data of magnitude $\varrho$ (in kW) under the three attack patterns of the threat model: a single injection on the masked variable $p_{i,t}^{dc}+w_{r,i}$, a joint compensating injection on both $p_{i,t}^{dc}+w_{r,i}$ and $\gamma_{i,t}+w_{r,i}$, and a ciphertext contamination that adds $Enc(PK_u,\varrho)$ to the transmitted ciphertext. For each pattern and magnitude, the detection rate is measured over 50 randomized trials, as reported in Table \ref{tab:fdi}. All injections whose propagated effect exceeds $\psi$ are detected with rate $1.0$. The single and joint injections are fully detected for $\varrho\ge1$~kW (and the single injection already from $\varrho=0.1$~kW, since its residual scales as $\pi\varrho$ with $\pi\ge10$), while the ciphertext contamination, whose residual scales as $\varrho$, is fully detected for $\varrho\ge1$~kW. Injections below the noise floor ($\varrho\le10^{-2}$~kW) are not flagged, but such perturbations are below $0.12\%$ of even the smallest shared value and thus have negligible operational impact. These results confirm the integrity guarantee of Proposition 1 for injections above the tolerance $\psi$.

\begin{table}[!t]
	\centering
	\caption{Detection Rate (\%) versus False Data Injection Magnitude $\varrho$ (kW)}
	\label{tab:fdi}
	\scriptsize
	\setlength{\tabcolsep}{4pt}
	\begin{tabular}{lccccccc}
		\toprule
		Attack pattern & $10^{-4}$ & $10^{-3}$ & $10^{-2}$ & $10^{-1}$ & $1$ & $10$ & $10^{2}$ \\
		\midrule
		Single injection & 0 & 0 & 0 & 100 & 100 & 100 & 100 \\
		Joint injection & 0 & 0 & 0 & 86 & 100 & 100 & 100 \\
		Ciphertext contamination & 0 & 0 & 0 & 0 & 100 & 100 & 100 \\
		\bottomrule
	\end{tabular}
	\vspace{-0.3cm}
\end{table}

Beyond detection, we quantify the operational impact of an injection on the dispatch. A compromised data center perturbs its shared power $p_{i,t}^{\text{dc}}$ by $\varrho$, and the utility solves $\mathcal{P}_3$ against the reported demand; the resulting dispatch cost distortion and physical power imbalance are reported in Fig. \ref{fig:poison_opt}. Without verification, the impact grows unboundedly with $\varrho$. A single-data-center injection sustained over the horizon inflates the dispatch cost by up to $8.6\%$ and induces a power imbalance of up to $2.4\times10^{4}$~kWh at $\varrho=10^{3}$~kW. With the verifiable double aggregation, every injection with $\varrho\ge\psi=1$~kW is detected and the contaminated share is rejected, eliminating the impact; the residual undetectable manipulation ($\varrho<1$~kW) is bounded below $0.01\%$ cost distortion and $24$~kWh imbalance, which is negligible for the system. Hence the verification mechanism not only detects but also strictly bounds the operational impact of data-integrity attacks.

\subsubsection{Robustness under model poisoning}
\begin{figure}[!hbpt]
	\centering
	\includegraphics[width=0.92\linewidth]{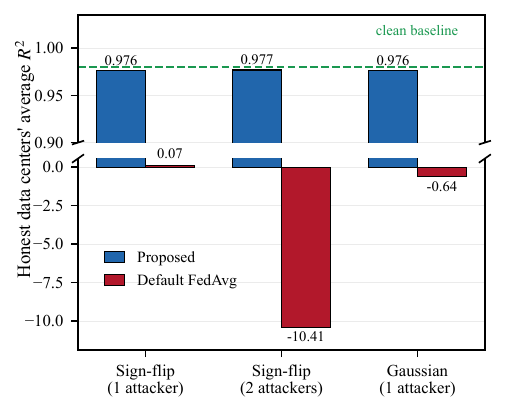}
	\caption{Honest data centers' average $R^2$ under model-poisoning attacks: the proposed adaptive federated learning versus default FedAvg.}
	\label{fig:poison_learn}
\end{figure}

We next evaluate the robustness of the adaptive federated learning against malicious data centers that submit poisoned model updates during training. Following the threat model, up to two data centers act as attackers that, at every aggregation round, corrupt their shared model parameter, $\theta$, by either a scaled sign-flip ($-4\theta$) or additive Gaussian noise ($\theta+\mathcal{N}(0,I)$). Under the training configuration of Section IV-A, we compare the proposed approach, in which each data center admits an aggregate update only if it passes the acceptance criterion, against default federated averaging (FedAvg) that unconditionally applies the aggregate. Table \ref{tab:poison} and Fig. \ref{fig:poison_learn} report the coefficient of determination $R^2$ averaged over the honest data centers.

\begin{figure}[!hbpt]
	\centering
	\includegraphics[width=0.92\linewidth]{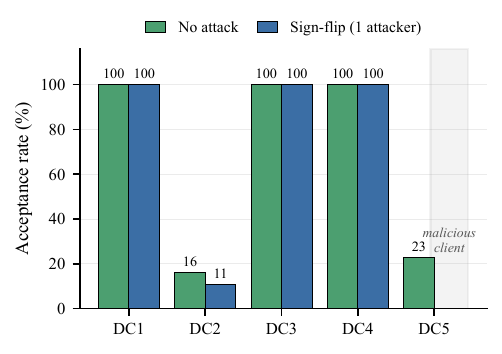}
	\caption{Per-data-center acceptance rate without and with a sign-flipping attacker; the heterogeneous acceptance pattern is preserved under attack.}
	\label{fig:poison_acc}
\end{figure}

Without a defense, a single sign-flipping attacker collapses the shared model. The honest data centers' average $R^2$ falls from the clean baseline of $0.980$ to $0.070$, and two attackers drive it to $-10.41$ (worse than predicting the mean), with the worst data center reaching $R^2=-25.30$; the Gaussian attacker is likewise destructive ($R^2=-0.64$). In contrast, the proposed approach preserves the honest performance almost intact, i.e., the honest average $R^2$ remains $0.976$, $0.977$, and $0.976$ under one sign-flipping, two sign-flipping, and one Gaussian attacker, respectively, within $0.004$ of the clean baseline. Fig. \ref{fig:poison_acc} further shows that the acceptance rates of the honest data centers under attack remain consistent with the attack-free case (data centers 1, 3, and 4 near $100\%$, data center 2 around $10$-$16\%$). The acceptance criterion filters out the contaminated aggregate updates while continuing to admit the beneficial ones, so the honest data centers retain a lower-bounded level of performance. These results substantiate the robustness goal of the threat model.

\begin{table}[!hbpt]
	\centering
	\caption{Honest Data Centers' $R^2$ Under Model Poisoning}
	\label{tab:poison}
	\scriptsize
	\setlength{\tabcolsep}{5pt}
	\begin{tabular}{llcc}
		\toprule
		Attack & Defense & Average $R^2$ & Minimum $R^2$ \\
		\midrule
		No attack & Proposed & 0.980 & 0.951 \\
		Sign-flip (1 attacker) & Proposed & 0.976 & 0.945 \\
		Sign-flip (1 attacker) & FedAvg & 0.070 & $-0.880$ \\
		Sign-flip (2 attackers) & Proposed & 0.977 & 0.954 \\
		Sign-flip (2 attackers) & FedAvg & $-10.41$ & $-25.30$ \\
		Gaussian (1 attacker) & Proposed & 0.976 & 0.952 \\
		Gaussian (1 attacker) & FedAvg & $-0.640$ & $-3.148$ \\
		\bottomrule
	\end{tabular}
	\vspace{-0.3cm}
\end{table}

\section{Conclusions}
Data centers function as multi-energy systems that integrate electricity, heat, and data flows. The optimization of such integrated multi-domain flows often involves mixed-integer formulations and access to proprietary or sensitive datasets, which can correspondingly introduce computational burdens and raise data privacy concerns. To address these challenges, this paper proposes an adaptive federated learning-to-optimization approach that accounts for the heterogeneity of datasets across geographically distributed data centers. To safeguard data privacy, secret sharing techniques are incorporated into both the learning and optimization stages. Furthermore, a model acceptance criterion with convergence guarantee is developed in the learning stage to enhance training performance while filtering out potentially contaminated data. In the optimization stage, a verifiable double aggregation mechanism is proposed to simultaneously ensure the privacy and integrity of shared data. Theoretical analysis and numerical simulations verify that the proposed adaptive federated learning-to-optimization approach: i) can simultaneously ensure the privacy and integrity of the shared variables in the data center optimization; ii) ensures the near-optimality; iii) exhibits good computational efficiency for the data center optimization.  
 
{\appendix
	
\subsection{Preliminaries of the Adaptive Federated Learning}

At learning round $r$, agent $i$, $i \in \mathcal{I}_r$ ($|\mathcal{I}_r|=m$), produces local models $\theta_{r,i}$ from the same reference $\bar\theta_r$. The average parameters from the aggregator are
$\bar\theta_r=\sum_{i\in\mathcal{I}_r} \overline{w}_{r,i}\theta_{r,i}$ with $\sum_i \overline{w}_{r,i}=1$, $\overline{w}_{r,i}={w}_{r,i}/{w}_{r}$, and
$0<\gamma_{\min}\le \overline{w}_{r,i}\le \gamma_{\max}<\infty$. Let $f_i:\mathbb{R}^d\!\to\!\mathbb{R}$ be convex and $L$-smooth, and
$f:=\tfrac1m\sum_{i=1}^m f_i$, $f^\star:=\inf_x f(x)$.
We can have:
\begin{align}
&v_{r,i}=\beta\,v_{r-1,i}+(1-\beta)\,(\theta_{r,i}-\bar\theta_r),\quad \beta\in[0,1), \label{eq:ema}\\
&\theta_{r+1,i}=\theta_{r,i}+\alpha_r\,C_{r,i}\,v_{r,i},\quad C_{r,i}\in[0,1]. \label{eq:mix}
\end{align}
Define the effective gains $\lambda_{r,i}=\alpha_r(1-\beta)C_{r,i}$ and
$\tau_r=\sum_i \overline{w}_{r,i}\lambda_{r,i}$. We assume $0<\tau_{\min}\le \tau_r\le \tau_{\max}$. We can define simpler symbols as $V_r=\frac1{m}\sum_{i}\|v_{r,i}\|^2$ and also:
\begin{align}
&\Delta_r=\E[f(\bar\theta_r)-f^\star],\quad
D_r=\frac1{m}\sum_{i}\|\theta_{r,i}-\bar\theta_r\|^2. \label{ap_1}
\end{align}
Let $\tilde g_{r}=\sum_i\overline{w}_{r,i}\tilde g_{r,i}$, where $\tilde g_{r,i}$ is the global-unbiased gradient. Let $B_{r,i}$ be the size of the local mini-batch for agent $i$. The global-unbiased control-variates \cite{karimireddy2020scaffold, stich2018local, acar2021federated} are summarized as:
\begin{subequations}
\begin{align}
&\theta_{r,i}=\bar\theta_r-\eta_r^{\rm loc}\,\tilde g_{r,i},\qquad
\E[\tilde g_{r,i}\mid \bar\theta_r]=\nabla f(\bar\theta_r)=g_r, \label{eq:sur}\\
&\E\|\tilde g_{r,i}-\nabla f(\bar\theta_r)\|^2\le \frac{\sigma_i^2}{B_{r,i}}+\zeta^2,\\ &\sum_{i}\E(\overline{w}_{r,i}\|\tilde g_{r,i}-\nabla f(\bar\theta_r)\|^2)\le \frac{\sigma^2}{B_{r}}+\zeta^2.
\end{align}
\end{subequations}
Based on \eqref{ap_1}, we can further obtain:
\begin{subequations}
\begin{align}
D_r = \frac{(\eta_r^{loc})^2}{m}\sum_i\|\tilde g_{r,i}-\tilde g_{r}\|^2 \le \frac{(\eta_r^{loc})^2}{m\gamma_{min}}\sum_i (\overline{w}_{r,i}\|\tilde g_{r,i}- g_{r}\|^2)
\end{align}
\end{subequations}
For all inputs, $x$, the heterogeneity \cite{li2020federated} is maintained as:
\begin{equation}
\frac1m\sum_{i}\|\nabla f_i(x)-\nabla f(x)\|^2\le \zeta^2,
\label{eq:het}
\end{equation}
which implies $\sum_i \overline{w}_{r,i}\|\nabla f_i-\nabla f\|^2\le \zeta^2$ since $\sum_i \overline{w}_{r,i}=1$.

\begin{lemma}[Smoothness descent \cite{boyd2004convex}]
For $L$-smooth $f$ and $d_r=\bar\theta_{r+1}-\bar\theta_r$, it can be obtained:
\begin{equation}
\Delta_{r+1}\le \Delta_r+\E\langle \nabla f(\bar\theta_r),d_r\rangle+\tfrac{L}{2}\E\|d_r\|^2. 
\label{eq:smooth}
\end{equation}
\end{lemma}

\begin{lemma}[Exponential Moving Average energy recursion \cite{yu2019linear}]\label{lem:ema}
For \eqref{eq:ema}, there exist $a\in(0,1)$ and $A>0$ such that
\begin{equation}
V_r\ \le\ a\,V_{r-1}\ +\ A\,D_r.
\label{eq:ema-energy}
\end{equation}
\end{lemma}
\subsection{Convergence for the Adaptive Federated Learning}
Define $d_r=\bar\theta_{r+1}-\bar\theta_r$. From \eqref{eq:ema}--\eqref{eq:mix}, we can obtain:
\begin{equation}
d_r=\underbrace{\sum_i \overline{w}_{r,i}\lambda_{r,i}(\theta_{r,i}-\bar\theta_r)}_{U^{\text{a}}_r}
+\underbrace{\sum_i \overline{w}_{r,i}\beta\alpha_r C_{r,i} v_{r-1,i}}_{U^{\text{b}}_r}.
\end{equation}

By \eqref{eq:sur}, $\theta_{r,i}-\bar\theta_r=-\eta_r^{\rm loc}\tilde g_{r,i}$, hence we can obtain:
\begin{equation}
U^{\text{a}}_r
=-\,\eta_r^{\rm loc}\sum_i \overline{w}_{r,i}\lambda_{r,i}\,\tilde g_{r,i},
\quad
\E[\tilde g_{r,i}\mid\bar\theta_r]=\nabla f_i(\bar\theta_r).
\end{equation}
Conditioning on $\bar\theta_r$ and writing $g_r=\nabla f(\bar\theta_r)$, we can derive:
\begin{align}
\E\!\left[\langle g_r,U^{\text{a}}_r\rangle\mid \bar\theta_r\right]
&= -\,\eta_r^{\rm loc}\sum_i \overline{w}_{r,i}\lambda_{r,i}\,\langle g_r,\nabla f_i(\bar\theta_r)\rangle \nonumber\\
&\le -\,\eta_r^{\rm loc}\left(\frac{\tau_r}{2}\|g_r\|^2\right),
\label{eq:fresh-ip}
\end{align}
using the weighted identity $\sum_i \tilde{w}_{r,i}\langle g,\nabla f_i\rangle
=\|g\|^2+\sum_i \tilde{w}_{r,i}\langle g,\nabla f_i-\nabla f\rangle
\ge \tfrac12\|g\|^2 - \tfrac12\sum_i \tilde{w}_{r,i}\|\nabla f_i-\nabla f\|^2$, $\tilde{w}_{r,i}=\overline{w}_{r,i}\lambda_{r,i}/\sum_i \overline{w}_{r,i}\lambda_{r,i}$, and \eqref{eq:het}.

We can obtain the upper bound of inner product for $U^{\text{b}}_r$ using Young's inequality and Cauchy–Schwarz inequality. For any $\rho>0$, we can obtain:
\begin{equation}
\big|\E\langle g_r,U^{\text{b}}_r\rangle\big|
\le \frac{\rho}{2}\,\E\|g_r\|^2+\frac{\beta^2\alpha_r^2\gamma_{\max}}{2\rho}\,m\,\E V_{r-1}.
\label{eq:leg-ip}
\end{equation}

By employing the Jensen inequality and $\overline{w}_{r,i}\le\gamma_{\max}$, we can derive the upper bounds for the quadratic terms as:
\begin{equation}
\E\|U^{\text{a}}_r\|^2\le \gamma_{\max}\lambda_{\max}^2\,m\,D_r,
\quad
\E\|U^{\text{b}}_r\|^2\le \beta^2\alpha_r^2\gamma_{\max}\,m\,\E V_{r-1},
\label{eq:quads}
\end{equation}
where $\lambda_{\max}=\alpha_r(1-\beta)$ since $C_{r,i}\in[0,1]$.

Plug \eqref{eq:fresh-ip}, \eqref{eq:leg-ip}, \eqref{eq:quads} into \eqref{eq:smooth} and choose $\rho:=\frac{\eta_r^{\rm loc}\tau_r}{2}$ and $\E\|d_r\|^2 \le 2\E\|U^{\text{a}}_r\|^2+2\E\|U^{\text{b}}_r\|^2$. We can obtain the bound:
\begin{align}
\Delta_{r+1} & \le \Delta_r \notag \\
&-c_r\E\|\nabla f(\bar\theta_r)\|^2+C_DD_r+C_V\alpha_r^2\,m\,\E V_{r-1},
\label{eq:one-step}
\end{align}
for constants $c_r=\frac{\eta_r^{\rm loc}\tau_r}{4}$, $C_D=L\gamma_{\max}\lambda_{\max}^2 m$,
$C_V=\tfrac{\beta^2\gamma_{\max}}{\eta_r^{\rm loc}\tau_r}+L\beta^2\gamma_{\max}$
(depending on $L,\beta,\gamma_{\max}, \tau_r$).

Define the potential function as $\Psi_r=\Delta_r+\mu D_r+\nu V_r$ with some $\mu,\nu>0$.
First, \eqref{eq:ema-energy} gives $V_r\le aV_{r-1}+A D_r$.
The variance update in \eqref{eq:mix} implies
\begin{equation}
\E[D_{r+1}\mid\mathcal F_r]\ \le\ (1+c_\lambda)\,D_r + c_v\,\alpha_r^2\,\E V_{r-1},
\label{eq:D-rec}
\end{equation}
which can be obtained by squaring
$\theta_{r+1,i}-\bar\theta_{r+1}=\theta_{r,i}-\bar\theta_{r}+\alpha_r\,C_{r,i}\,v_{r,i}-\alpha_r\,\overline{C_{r}\,v_{r}}$ and using Young’s inequality, where $c_\lambda, c_v>0$ are constants and $\mathcal{F}_r$ denotes the filtration up to round $r$.
Combining \eqref{eq:one-step}, \eqref{eq:D-rec}, and $V_r\le aV_{r-1}+A D_r$, we can pick $\mu$ and $\nu$ so that the coefficients of $D_r$ and $V_{r-1}$ in $\Psi_{r+1}-\Psi_r$ are nonpositive. Hence, for some $\tilde C>0$, we can have:
\begin{equation}
\Psi_{r+1}-\Psi_r
\le\ -c_r\E\|\nabla f(\bar\theta_r)\|^2\ +\ \tilde C\,(\eta_r^{\rm loc})^2\,(\zeta^2+\frac{\sigma^2}{B_r}),
\label{eq:psi-drift}
\end{equation}
By the gradient assumption from the Polyak--\L{}ojasiewicz condition for $f$ \cite{karimi2016linear}, i.e., $\|g_r\|^2\ge 2 \mu_f\, \Delta_r$ with the constant $\mu_f>0$, we can further obtain:
\begin{equation}
\Psi_{r+1}-\Psi_r
\ \le\ -\,2\mu_f\,c_r\,\Delta_r\ +\ \tilde C\,(\eta_r^{\rm loc})^2\,(\zeta^2+\frac{\sigma^2}{B_r}).
\label{eq:psi-drift-gap}
\end{equation}
Sum \eqref{eq:psi-drift-gap} for $r=0,\dots,T-1$ and use $\Psi_T\ge 0, \Delta_r\ge0$ for the loss minimization problem:
\begin{equation}
2\mu_f\sum_{r=0}^{T-1} c_r\,\Delta_r
\ \le\ \Psi_0+\tilde C\,(\zeta^2+\frac{\sigma^2}{B_r})\sum_{r=0}^{T-1}(\eta_r^{\rm loc})^2.
\end{equation}
Let $W_T=\sum_{r=0}^{T-1} c_r$ and define the $c_r$-weighted output
$\hat\theta_T=\sum_{r=0}^{T-1}\alpha_r^{\rm out}\bar\theta_r$, with
$\alpha_r^{\rm out}=c_r/W_T$.
By convexity and $\Delta_r=\E[f(\bar\theta_r)-f^\star]$, we can derive the relation as:
\begin{align}
\E[f(\hat\theta_T)-f^\star]
&\le \frac{\sum_r c_r\Delta_r}{W_T} \nonumber\\
\le& \frac{\Psi_0}{2\mu_f\,W_T} + \frac{\tilde C}{2\mu_f}\cdot \frac{\sum_r (\eta_r^{\rm loc})^2}{W_T}\,(\zeta^2+\frac{\sigma^2}{B_r}).
\end{align}
If $\eta_r^{\rm loc}=\eta_0/\sqrt{r+1}$ and $\tau_r\ge \tau_{\min}>0$, then
$c_r=\tfrac14\eta_r^{\rm loc}\tau_r\ge \tfrac14\eta_0\tau_{\min}/\sqrt{r+1}$, so
$W_T=\Theta(\sqrt{T})$ and $\sum_r \eta_r^{\rm loc}=\Theta(\sqrt{T})$. Let $C_1$ and $C_2$ be constants. Therefore we can obtain:
\begin{subequations}
\begin{align}
&\E[f(\hat\theta_T)-f^\star]\ \le\ \frac{C_1}{\sqrt{T}}+\frac{C_2\log T}{\sqrt{T}}\,(\zeta^2+\sigma^2).
\end{align}
\end{subequations}
It follows that the adaptive federated learning converges to the optimal value, i.e., $\E[f(\hat\theta_T)-f^\star] \rightarrow 0$ as $T \rightarrow \infty$. \qed

\bibliographystyle{IEEEtran}
\bibliography{index}

\begin{thebibliography}{10}
\providecommand{\url}[1]{#1}
\csname url@samestyle\endcsname
\providecommand{\newblock}{\relax}
\providecommand{\bibinfo}[2]{#2}
\providecommand{\BIBentrySTDinterwordspacing}{\spaceskip=0pt\relax}
\providecommand{\BIBentryALTinterwordstretchfactor}{4}
\providecommand{\BIBentryALTinterwordspacing}{\spaceskip=\fontdimen2\font plus
\BIBentryALTinterwordstretchfactor\fontdimen3\font minus
  \fontdimen4\font\relax}
\providecommand{\BIBforeignlanguage}[2]{{%
\expandafter\ifx\csname l@#1\endcsname\relax
\typeout{** WARNING: IEEEtran.bst: No hyphenation pattern has been}%
\typeout{** loaded for the language `#1'. Using the pattern for}%
\typeout{** the default language instead.}%
\else
\language=\csname l@#1\endcsname
\fi
#2}}
\providecommand{\BIBdecl}{\relax}
\BIBdecl

\bibitem{zhang2023research}
Y.~Zhang, K.~Shan, X.~Li, H.~Li, and S.~Wang, ``Research and technologies for
  next-generation high-temperature data centers--state-of-the-arts and future
  perspectives,'' \emph{Renewable and Sustainable Energy Reviews}, vol. 171, p.
  112991, 2023.

\bibitem{koot2021usage}
M.~Koot and F.~Wijnhoven, ``Usage impact on data center electricity needs: A
  system dynamic forecasting model,'' \emph{Applied Energy}, vol. 291, p.
  116798, 2021.

\bibitem{fan2025highly}
W.~Fan, Y.~Pan, F.~Xiao, P.~Zhang, L.~Han, and S.-Y. Hsieh, ``A highly scalable
  network architecture for optical data centers,'' \emph{IEEE Transactions on
  Computers}, 2025.

\bibitem{yin2022exploiting}
X.~Yin, C.~Ye, Y.~Ding, and Y.~Song, ``Exploiting internet data centers as
  energy prosumers in integrated electricity-heat system,'' \emph{IEEE
  Transactions on Smart Grid}, vol.~14, no.~1, pp. 167--182, 2022.

\bibitem{rong2016optimizing}
H.~Rong, H.~Zhang, S.~Xiao, C.~Li, and C.~Hu, ``Optimizing energy consumption
  for data centers,'' \emph{Renewable and Sustainable Energy Reviews}, vol.~58,
  pp. 674--691, 2016.

\bibitem{long2023collaborative}
X.~Long, Y.~Li, Y.~Li, L.~Ge, H.~B. Gooi, C.~Chung, and Z.~Zeng,
  ``Collaborative response of data center coupled with hydrogen storage system
  for renewable energy absorption,'' \emph{IEEE Transactions on Sustainable
  Energy}, vol.~15, no.~2, pp. 986--1000, 2023.

\bibitem{11142335}
J.~Liu, F.~Teng, and Y.~Hou, ``Synergising hierarchical data centers and power
  networks: A privacy-preserving approach,'' \emph{IEEE Transactions on Smart
  Grid}, pp. 1--1, 2025.

\bibitem{faquir2021cybersecurity}
D.~Faquir, N.~Chouliaras, V.~Sofia, K.~Olga, and L.~Maglaras, ``Cybersecurity
  in smart grids, challenges and solutions,'' \emph{AIMS Electronics and
  Electrical Engineering}, vol.~5, no.~1, pp. 24--37, 2021.

\bibitem{karale2021challenges}
A.~Karale, ``The challenges of iot addressing security, ethics, privacy, and
  laws,'' \emph{Internet of Things}, vol.~15, p. 100420, 2021.

\bibitem{saraswat2022protecting}
A.~K. Saraswat and V.~Meel, ``Protecting data in the 21st century: Challenges,
  strategies and future prospects,'' \emph{Information technology in industry},
  vol.~10, no.~2, pp. 26--35, 2022.

\bibitem{aminifar2024privacy}
A.~Aminifar, M.~Shokri, and A.~Aminifar, ``Privacy-preserving edge federated
  learning for intelligent mobile-health systems,'' \emph{Future Generation
  Computer Systems}, vol. 161, pp. 625--637, 2024.

\bibitem{alhazmi2025federated}
M.~Alhazmi, A.~P. Zhao, W.~Li, and C.~Yang, ``Federated learning for real-time
  demand response by data centers toward energy efficiency and privacy
  preservation,'' \emph{IEEE Access}, 2025.

\bibitem{yang2022federated}
Z.~Yang, M.~Chen, K.-K. Wong, H.~V. Poor, and S.~Cui, ``Federated learning for
  6g: Applications, challenges, and opportunities,'' \emph{Engineering},
  vol.~8, pp. 33--41, 2022.

\bibitem{ahmad2024lightweight}
M.~S. Ahmad and S.~M. Shah, ``A lightweight mini-batch federated learning
  approach for attack detection in iot,'' \emph{Internet of Things}, vol.~25,
  p. 101088, 2024.

\bibitem{mantey2024federated}
E.~A. Mantey, C.~Zhou, J.~H. Anajemba, J.~K. Arthur, Y.~Hamid, A.~Chowhan, and
  O.~O. Otuu, ``Federated learning approach for secured medical recommendation
  in internet of medical things using homomorphic encryption,'' \emph{IEEE
  Journal of Biomedical and Health Informatics}, vol.~28, no.~6, pp.
  3329--3340, 2024.

\bibitem{rieyan2024advanced}
S.~A. Rieyan, M.~R.~K. News, A.~M. Rahman, S.~A. Khan, S.~T.~J. Zaarif,
  M.~G.~R. Alam, M.~M. Hassan, M.~Ianni, and G.~Fortino, ``An advanced data
  fabric architecture leveraging homomorphic encryption and federated
  learning,'' \emph{Information Fusion}, vol. 102, p. 102004, 2024.

\bibitem{saad2025towards}
Z.~Saad, J.~Yang, H.~Leung, and S.~Drew, ``Towards carbon-aware container
  orchestration: Predicting workload energy consumption with federated
  learning,'' \emph{arXiv preprint arXiv:2510.03970}, 2025.

\bibitem{liu2023privacy}
J.~Liu, Q.~Long, R.-P. Liu, W.~Liu, X.~Cui, and Y.~Hou, ``Privacy-preserving
  peer-to-peer energy trading via hybrid secure computations,'' \emph{IEEE
  Transactions on Smart Grid}, vol.~15, no.~2, pp. 1951--1964, 2023.

\bibitem{bonawitz2017practical}
K.~Bonawitz, V.~Ivanov, B.~Kreuter, A.~Marcedone, H.~B. McMahan, S.~Patel,
  D.~Ramage, A.~Segal, and K.~Seth, ``Practical secure aggregation for
  privacy-preserving machine learning,'' in \emph{Proceedings of the 2017 ACM
  SIGSAC Conference on Computer and Communications Security}, 2017, pp.
  1175--1191.

\bibitem{zhang2020batchcrypt}
C.~Zhang, S.~Li, J.~Xia, W.~Wang, F.~Yan, and Y.~Liu, ``Batchcrypt: Efficient
  homomorphic encryption for cross-silo federated learning,'' in \emph{2020
  USENIX Annual Technical Conference (USENIX ATC 20)}, 2020, pp. 493--506.

\bibitem{xu2020verifynet}
G.~Xu, H.~Li, S.~Liu, K.~Yang, and X.~Lin, ``Verifynet: Secure and verifiable
  federated learning,'' \emph{IEEE Transactions on Information Forensics and
  Security}, vol.~15, pp. 911--926, 2020.

\bibitem{guo2021verifl}
X.~Guo, Z.~Liu, J.~Li, J.~Gao, B.~Hou, C.~Dong, and T.~Baker, ``Verifl:
  Communication-efficient and fast verifiable aggregation for federated
  learning,'' \emph{IEEE Transactions on Information Forensics and Security},
  vol.~16, pp. 1736--1751, 2021.

\bibitem{li2020federated}
T.~Li, A.~K. Sahu, M.~Zaheer, M.~Sanjabi, A.~Talwalkar, and V.~Smith,
  ``Federated optimization in heterogeneous networks,'' \emph{Proceedings of
  Machine learning and systems}, vol.~2, pp. 429--450, 2020.

\bibitem{karimireddy2020scaffold}
S.~P. Karimireddy, S.~Kale, M.~Mohri, S.~Reddi, S.~Stich, and A.~T. Suresh,
  ``Scaffold: Stochastic controlled averaging for federated learning,'' in
  \emph{International conference on machine learning}.\hskip 1em plus 0.5em
  minus 0.4em\relax PMLR, 2020, pp. 5132--5143.

\bibitem{benaissa2021tenseal}
A.~Benaissa, B.~Retiat, B.~Cebere, and A.~E. Belfedhal, ``Tenseal: A library
  for encrypted tensor operations using homomorphic encryption,'' \emph{arXiv
  preprint arXiv:2104.03152}, 2021.

\bibitem{liu2025byzantine}
J.~Liu, Q.~Long, R.-P. Liu, W.~Liu, and Y.~Hou, ``Byzantine-resilient
  distributed p2p energy trading via spatial-temporal anomaly detection,''
  \emph{IEEE Transactions on Smart Grid}, 2025.

\bibitem{singh2024blockchain}
N.~Singh, H.~P. Singh, A.~Mishra, A.~Khare, M.~Swarnkar, and S.~K. Almas,
  ``Blockchain cloud computing: comparative study on ddos, mitm and sql
  injection attack,'' in \emph{2024 IEEE International Conference on Big Data
  \& Machine Learning (ICBDML)}.\hskip 1em plus 0.5em minus 0.4em\relax IEEE,
  2024, pp. 73--78.

\bibitem{gurobi}
\BIBentryALTinterwordspacing
{Gurobi Optimization, LLC}, ``{Gurobi Optimizer Reference Manual},'' 2024.
  [Online]. Available: \url{https://www.gurobi.com}
\BIBentrySTDinterwordspacing

\bibitem{junhong2025}
J.~Liu, R.-P. Liu, Y.~Hou \emph{et~al.}, ``data-center-dataset,''
  \url{https://github.com/johnny-eee/data-center-dataset}, 2025.

\bibitem{stich2018local}
S.~U. Stich, ``Local sgd converges fast and communicates little,'' \emph{arXiv
  preprint arXiv:1805.09767}, 2018.

\bibitem{acar2021federated}
D.~A.~E. Acar, Y.~Zhao, R.~M. Navarro, M.~Mattina, P.~N. Whatmough, and
  V.~Saligrama, ``Federated learning based on dynamic regularization,''
  \emph{arXiv preprint arXiv:2111.04263}, 2021.

\bibitem{boyd2004convex}
S.~P. Boyd and L.~Vandenberghe, \emph{Convex optimization}.\hskip 1em plus
  0.5em minus 0.4em\relax Cambridge university press, 2004.

\bibitem{yu2019linear}
H.~Yu, R.~Jin, and S.~Yang, ``On the linear speedup analysis of communication
  efficient momentum sgd for distributed non-convex optimization,'' in
  \emph{International Conference on Machine Learning}.\hskip 1em plus 0.5em
  minus 0.4em\relax PMLR, 2019, pp. 7184--7193.

\bibitem{karimi2016linear}
H.~Karimi, J.~Nutini, and M.~Schmidt, ``Linear convergence of gradient and
  proximal-gradient methods under the polyak-{\l}ojasiewicz condition,'' in
  \emph{Joint European conference on machine learning and knowledge discovery
  in databases}.\hskip 1em plus 0.5em minus 0.4em\relax Springer, 2016, pp.
  795--811.

\end{thebibliography}
\end{document}